\pdfoutput=1

\documentclass[aps,prd,twocolumn,superscriptaddress,preprintnumbers,nofootinbib]{revtex4-2}

\usepackage{amsmath}
\usepackage{amssymb}
\usepackage{hyperref}
\usepackage{tikz}
\usepackage[utf8]{inputenc}
\usetikzlibrary{decorations.markings}

\usepackage{enumitem}
\setlist[itemize]{align=parleft,left=10pt}

\let\imath\undefined
\newcommand{\imath}{\mathrm{i}}

\def\Section#1{\vskip .2 cm \noindent \textbf{#1}}

\begin{document}

\preprint{CALT-TH-2021-003, FR-PHENO-2021-02, OUTP-21-02P}

\title{Gravitational Bremsstrahlung from Reverse Unitarity}

\author{Enrico Herrmann}
\affiliation{Mani L. Bhaumik Institute for Theoretical Physics,\\
UCLA Department of Physics and Astronomy, Los Angeles, CA 90095, USA}

\author{Julio~Parra-Martinez}
\affiliation{Walter Burke Institute for Theoretical Physics, \\
California Institute of Technology, Pasadena, CA 91125, USA}

\author{Michael S. Ruf}
\affiliation{Physikalisches Institut, Albert-Ludwigs Universität Freiburg, D-79104 Freiburg, Germany}

\author{Mao Zeng}
\affiliation{Rudolf Peierls Centre for Theoretical Physics, \\
University of Oxford, Parks Road, Oxford OX1 3PU, United Kingdom
}

\begin{abstract}
%
We compute the total radiated momentum carried by gravitational waves during the scattering of two spinless black holes at the lowest order in Newton's constant, $\mathcal O(G^3)$, and all orders in velocity. By analytic continuation into the bound state regime, we obtain the ${\cal O}(G^3)$ energy loss in elliptic orbits. This provides an essential step towards the complete understanding of the third-post-Minkowskian binary dynamics. We employ the formalism of Kosower, Maybee, and O'Connell (KMOC) which relates classical observables to quantum scattering amplitudes and derive the relevant integrands using generalized unitarity. The subsequent phase-space integrations are performed via the reverse unitarity method familiar from collider physics, using differential equations to obtain the exact velocity dependence from near-static boundary conditions. 
\end{abstract}

\maketitle

\Section{Introduction.}
\label{sec:intro}
%
There has been enormous progress in applying scattering amplitude tools, such as generalized unitarity \cite{Bern:1994zx,Bern:1994cg,Britto:2004nc} and the double copy \cite{Bern:2008qj,Bern:2010ue,Bern:2012uf,Bern:2017ucb, Bern:2018jmv,Bern:2019prr}, together with effective field theory ideas \cite{Goldberger:2004jt,Porto:2016pyg,Cheung:2018wkq}, to the classical relativistic two-body problem, geared towards applications for current and future gravitational wave detectors \cite{Abbott:2016blz,*TheLIGOScientific:2017qsa,Punturo_2010,*Audley:2017drz,*Reitze:2019iox}. Such techniques have produced new results for the dynamics of spinless \cite{Damour:2016gwp,Foffa:2016rgu,Blumlein:2019zku,Blumlein:2020znm,Bjerrum-Bohr:2018xdl,Cheung:2018wkq,Bern:2019nnu,Foffa:2019hrb,Bern:2019crd,Cheung:2020gyp,Kalin:2020mvi,Kalin:2020fhe,Cristofoli:2020uzm,Bini:2020uiq,Bini:2020rzn,Loebbert:2020aos,Bern:2021dqo} and spinning \cite{Vaidya:2014kza,Vines:2017hyw,Guevara:2017csg,Vines:2018gqi,Guevara:2018wpp,Chung:2018kqs,Guevara:2019fsj,Chung:2019duq,Damgaard:2019lfh,Aoude:2020onz,Bern:2020buy,Guevara:2020xjx,Levi:2020kvb,Levi:2020uwu} black holes, including finite-size effects \cite{Cheung:2020sdj,Haddad:2020que,Kalin:2020lmz,Brandhuber:2019qpg,Huber:2019ugz,AccettulliHuber:2020oou,AccettulliHuber:2020oou,Bern:2020uwk,Cheung:2020gbf,Aoude:2020ygw}.  This effort has mainly focused on the conservative dynamics, described by a two-body Hamiltonian~\cite{Cheung:2018wkq,Cristofoli:2019neg}, or the scattering gravitational waveform \cite{Goldberger:2016iau, Luna:2017dtq, Shen:2018ebu,Bautista:2019tdr,Mogull:2020sak,Huber:2020xny}. In this letter we use amplitude methods to compute a radiative observable for a bound binary system in general relativity. We do so by first calculating the momentum emitted in the form of gravitational waves during the scattering of two spinless black holes at ${\cal O}(G^3)$ and all orders in velocity. By analytic continuation from scattering to bound kinematics \cite{Kalin:2019rwq,Kalin:2019inp,Bini:2020hmy} we obtain the ${\cal O}(G^3)$ energy loss in an elliptic orbit of a binary system. 

We use Kosower, Maybee, and O'Connell's (KMOC) \cite{Kosower:2018adc} formalism to express classical observables directly in terms of scattering amplitudes and their unitarity cuts. Recently, this formalism has been used \cite{A:2020lub} to understand classical soft radiation \cite{Laddha:2018rle,Laddha:2018myi,Sahoo:2018lxl,Laddha:2019yaj,Saha:2019tub}. By focusing on an inclusive observable, involving a sum over final states of the scattering event, we avoid the need for detailed knowledge of the gravitational waveform and subtleties arising from infrared divergences in its phase. In a well defined sense, observables calculated in the KMOC formalism are analogous to inclusive cross sections in collider physics. Taking this analogy seriously allows us to import crucial technology developed in the particle-physics context. Concretely, we use generalized unitarity (based on the factorization of loop quantities into products of simpler on-shell tree-level amplitudes \cite{Bern:1994zx,Bern:1994cg,Britto:2004nc}) to construct the loop integrands; we employ (canonical) differential equations \cite{Kotikov:1990kg,Bern:1992em,Gehrmann:1999as,Henn:2013pwa, Henn:2014qga} adapted to the post-Minkowskian (PM) expansion, in powers of Newton's constant $G$, in classical gravity \cite{Parra-Martinez:2020dzs}, together with the reverse unitarity method \cite{Anastasiou:2002yz,Anastasiou:2002qz,Anastasiou:2003yy,Anastasiou:2015yha} for the phase-space integration.

There are alternative approaches to the radiation reaction problem in hyperbolic scattering \cite{Amati:1990xe, DiVecchia:2019myk, DiVecchia:2019kta, Bern:2020gjj, DiVecchia:2020ymx,  DiVecchia:2021ndb}. At 3PM order, the radiative effects on the scattering angle have been previously computed by amplitudes-based eikonal techniques in maximal supergravity \cite{DiVecchia:2020ymx} and by classical methods in general relativity \cite{Damour:2020tta}. With the radiated energy and momentum, our work provides a key missing ingredient of the full radiative 3PM dynamics. Furthermore, we compare to post-Newtonian (PN) results in the literature \cite{Kovacs:1978eu, Blanchet:1989cu,Bini:2020hmy,Peters:1963ux,Peters:1964zz,Wagoner:1976am,Blanchet:1989cu,Junker:1992kle,Gopakumar:1997ng,Gopakumar:2001dy,Arun:2007sg,Blanchet:2013haa}, which give this quantity to fixed orders in a small velocity expansion, finding agreement. 

\Section{KMOC formalism.}
\label{sec:KMOC}
%
In this work, we compute classical gravitational observables in the KMOC formalism \cite{Kosower:2018adc}. The basic idea of this approach is to set up a gedanken experiment for the scattering of two wavepackets, widely separated by an impact parameter $b^\mu$, and measure the change in an observable $O$, with corresponding quantum operator $\mathbb{O}$, between \emph{in} and \emph{out} states 
\begin{equation}
\label{eq:kmoc_start}
 \Delta O = \langle {\rm out} | \mathbb{O} | {\rm out} \rangle  - \langle {\rm in} | \mathbb{O} | {\rm in} \rangle\,.
\end{equation}
Since the \emph{out} state is related to the \emph{in} state via the time evolution operator, i.e.~the $S$-matrix, $| {\rm out} \rangle {=} S | {\rm in} \rangle$, we can write the change in the observable in terms of the scattering amplitude  $\mathcal{M} {=} -\imath (S{-}1)$, while still capturing the real time dynamics of the scattering process. For instance, the gravitational impulse $\Delta p_i^{\mu}$ is given as momentum difference of particle $i$ by measuring operator~$\mathbb{P}_i$ and the radiated momentum $\Delta R^\mu$ by measuring $\mathbb{R}^\mu$.

We are interested in \emph{classical} observables. This corresponds to the regime where the Compton wavelength of the particles representing the black holes is the smallest length scale in the problem (we point to Ref.~\cite{Kosower:2018adc} for a detailed discussion of this limit). For us, it suffices to state that we are interested in regions of external kinematics where the massive particle momenta $p_i$ scale like $\mathcal{O}(1)$ in the classical counting and the four-momentum transfer $q$, as well as the graviton loop variables $\ell_i$ scale like $\mathcal{O}(\hbar)$. Employing the terminology from the ``method of regions'' \cite{Beneke:1997zp}, the classical $\hbar$ expansion is then equivalent to the so-called soft expansion. This classical counting will play a crucial role when constructing loop integrands and evaluating the corresponding integrals.

In the classical limit, the dependence on the shape of the wavepackets drops out and one arrives at\,\footnote{We follow Ref.~\cite{Kosower:2018adc} and define $\hat{\mathrm{d}}x = \mathrm{d}x/(2\pi)$ and  $\hat\delta(x) = 2\pi \delta(x)$.}
\begin{equation}
\label{eq:KMOC_DObs}
    \Delta O = \int \hat{\mathrm{d}}^D q \,
    \hat\delta(-2 p_1 \cdot q)\, \hat\delta(2 p_2 \cdot q) e^{\imath b\cdot q}\,
    \left(
    \mathcal{I}_{O, {\rm v}} + \mathcal{I}_{O,{\rm r}}
    \right)
\end{equation}
where, borrowing language from collider observables, one can define \emph{virtual} and \emph{real}  kernels, $\mathcal{I}_{{\cal O}, {\rm v}}$ and  $\mathcal{I}_{{\cal O}, {\rm r}}$, which respectively depend on the virtual amplitude, and its unitarity cuts including a phase-space integration akin to those appearing in cross sections. The observable of interest is specified by a corresponding \emph{measurement function}. In the KMOC formalism, this amounts to a numerator insertion or differential operator acting on the component amplitudes in Eq.~(\ref{eq:KMOC_DObs}).

In this letter, we focus on the radiated momentum, $\Delta R^\mu$.~As explained in Ref.~\cite{Kosower:2018adc}, this observable only receives \emph{real} contributions, and the corresponding kernel expressed in terms of scattering amplitudes
\begin{align}
    \mathcal{I}_{R, {\rm r}}^\mu &= \sum_X 
    \int\!\mathrm{d}\Phi_{2+X}\  \ell^{\mu}_X \times \quad
    \raisebox{-32pt}{
    \includegraphics[scale=.8, trim = 35 0 0 0]{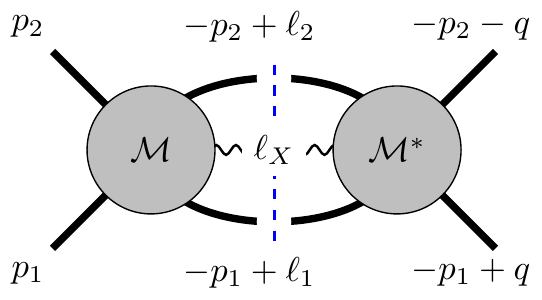}}
    \nonumber \\
    &= \sum_X \int\!\mathrm{d}\Phi_{2+X}(\ell_1,\ell_2,\{\ell_X\}) 
     \ \ell_X^\mu \label{eq:KMOC_kernel_real} \\
    &\phantom{=} \times 
    \mathcal{M}_{4{+}X}(p_1,p_2,{-}p_2{+}\ell_2,{-}p_1{+}\ell_1,\{\ell_X\})  \nonumber  \\
    &\phantom{=} \times \mathcal{M}^{\ast}_{4{+}X}(\{-\ell_X\},p_1{-}\ell_1,p_2{-}\ell_2,{-}p_2{-}q,{-}p_1{+}q)\,,\nonumber
\end{align}
is given by a sum over unitarity cuts featuring the exchange of sets of \emph{messengers}, $X$, in our case gravitons, including the empty set. As usual, such unitarity cuts involve an integral over the $n$-point Lorentz invariant phase space (d$\Phi_n$).  The measuring function for $\Delta R^\mu$ is encoded in the insertion of $\ell_X^\mu$, representing the total momentum carried by the messengers. We use an ``all-outgoing" convention for the momenta in the scattering amplitudes and mostly-minus signature metric. Eq.~\eqref{eq:KMOC_kernel_real} is closely related to the textbook formula for the radiated momentum \cite{Misner:1974qy}
\begin{equation}
    \Delta R^\mu = \int 
    \mathrm{d}\Phi_{1}(k)\, k^\mu\, h^*_{\nu\rho}(k)h^{\nu\rho}(k) \,,
\end{equation}
in terms of the momentum-space gravitational waveform, $h_{\mu\nu}(k)$. However, the calculation of the waveform requires computing multi-scale integrals depending on different components of $k^\mu$, whereas, as we will argue below, the direct computation of the multi-particle phase-space integral involves simpler functions of a single scale.

Although Eq.~\eqref{eq:KMOC_kernel_real} is valid  beyond perturbation theory, in this work, we will expand it perturbatively in $GE_{\rm cm}/b$, where $G$ is Newton's constant, and $E_{\rm cm}$ the center of mass energy. The first contribution to $\Delta R^\mu$ arises at ${\cal O}(G^3)$, since Bremsstrahlung of finite energy gravitons can only occur once one of the black holes is deflected due to its gravitational interaction with the other.
\begin{figure}[t]
	\centering
		\begin{minipage}{0.3\linewidth}
     	\centering
	    \includegraphics[height=1.5cm]{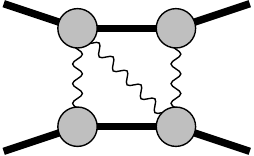}
        \end{minipage}
        \hspace{5pt}
		\begin{minipage}{0.3\linewidth}
     	\centering
	    \raisebox{0cm}{ \includegraphics[height=1.55cm]{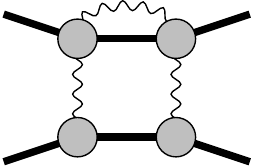}
	    }
        \end{minipage}
        \caption{Generalized unitarity cuts relevant for the radiated momentum. Shaded blobs denote tree-level amplitudes, visible legs are on shell, and we exclude any phase-space integrals.}
        \vspace{-6pt}
        \label{fig:spanningcuts}
\end{figure}

\Section{Integrands from generalized unitarity.}
\label{sec:integrands}
%
With an eye towards more general observables, at fixed order in $G$, instead of computing both real and virtual contributions in Eq.~(\ref{eq:KMOC_DObs}) separately, we obtain the integrand for the \emph{virtual} amplitudes and then take appropriate cuts and insert $\ell^\mu_X$ from Eq.~(\ref{eq:KMOC_kernel_real}) to obtain the \emph{real} contribution relevant for the radiated momentum. The virtual integrand for two-to-two scattering of massive scalars is derived by generalized unitarity \cite{Bern:1994zx,Bern:1994cg,Britto:2004nc} as described e.g., in Refs.~\cite{Bern:2019nnu,Bern:2019crd}. Unlike for the conservative integrand obtained in previous works \cite{Bern:2019nnu,Bern:2019crd}, in order to construct the scattering amplitude in the full soft region we are forced to include additional terms that are necessary, once radiation is taken into account. However, in order to capture all terms relevant for the radiated momentum in the classical limit, it suffices to match the set of cuts in Fig.~\ref{fig:spanningcuts}, the first of which is familiar from the conservative sector~\cite{Bern:2019nnu,Bern:2019crd}. Relative to the three-particle cut in Eq.~\eqref{eq:KMOC_kernel_real}, these have the advantage that only four-particle tree amplitudes are involved, and matter contact terms, which are quantum contributions, are automatically dropped. We compute these unitarity cuts by sewing tree-level amplitudes in $D=4{-}2\epsilon$ dimensions taking advantage of the additional simplifications of generalized gauge invariance \cite{KoemansCollado:2019ggb,Kosmopoulos:2020pcd}.

\begin{figure}[t]
	\centering
		\begin{minipage}{0.24\linewidth}
     	\centering
	    \includegraphics[height=0.75cm]{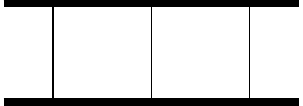}
        \end{minipage}
		\begin{minipage}{0.24\linewidth}
     	\centering
	    \includegraphics[height=0.75cm]{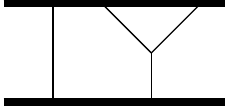}
        \end{minipage}
		\begin{minipage}{0.24\linewidth}
     	\centering
	    \includegraphics[height=0.75cm]{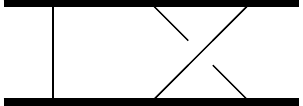}
        \end{minipage}
		\begin{minipage}{0.24\linewidth}
     	\centering
	    \scalebox{-1}[.9]{\includegraphics[height=0.85cm]{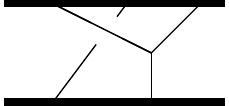}}
        \end{minipage}\\[7pt]
        \begin{minipage}[b]{0.24\linewidth}
     	\centering
     	\raisebox{0pt}{
	    \includegraphics[height=1cm]{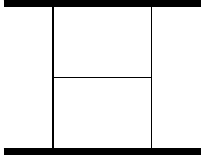}
	    }
        \end{minipage}
		\begin{minipage}[b]{0.24\linewidth}
     	\centering
	    \includegraphics[height=1cm]{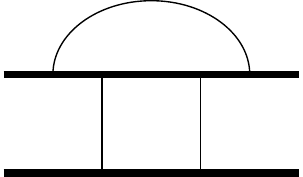}
        \end{minipage}
		\begin{minipage}[b]{0.24\linewidth}
     	\centering
     	\scalebox{-1}[1]{\includegraphics[height=0.85cm]{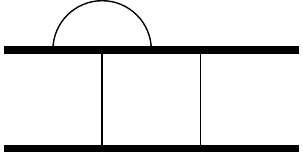}}
        \end{minipage}
		\begin{minipage}[b]{0.24\linewidth}
     	\centering
	    \includegraphics[height=0.85cm]{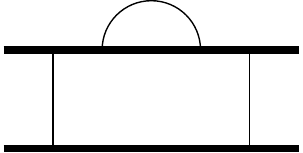}
        \end{minipage}
\caption{Cubic diagrams relevant for the radiated momentum.}
\vspace{-6pt}
\label{fig:all_diags}
\end{figure}

To find a diagrammatic representation of the integrand that matches the unitarity cuts in Fig.~\ref{fig:spanningcuts}, we write an ansatz in terms of the cubic diagrams in Fig.~\ref{fig:all_diags} with kinematic numerators. The first five graphs were already present in the conservative result, however, since we are now dealing with additional cuts, their numerators might slightly change. Furthermore, there are three new graphs which contribute to the radiated momentum computation. Each kinematic numerator of a graph $\Gamma$ in Fig.~\ref{fig:all_diags} is written as a polynomial of Lorentz products of the independent external momenta $p_i$ and the loop momenta $\ell_j$ up to mass dimension twelve. In addition, we impose the following constraints on the ansatz:
\begin{itemize}[itemsep=-1pt]
    \item apply on-shell conditions on all legs that are always cut in the spanning set of cuts depicted in Fig.~\ref{fig:spanningcuts},
    \item demand diagram symmetries of each graph (compatible with the cuts in the previous bullet point),
    \item a numerator of a diagram with $n$ all-graviton vertices is at least of order $|q|^{4 n}$ in the classical counting,
    \item the numerator of a diagram with iterated $s$-channel two-particle cuts containing $n_2$ tree-amplitudes is proportional to $n_2$ powers of the tree numerator.
\end{itemize}
The above rules fix e.g.~the numerator of the planar double box, that is, the first diagram in Fig.~\ref{fig:all_diags}, up to an overall normalization to $N_{\text{III}} {=} a_1 [(s{-}m_1^2{-}m^2_2)^2 {-} 4 m^2_1m^2_2]^3$ and leave 3731 parameters in the ansatz globally. Matching this ansatz against the cuts in Fig.~\ref{fig:spanningcuts} then determines the unknown parameters. All that is left to do is to evaluate the three-particle cut of the resulting integrand, perform the phase-space integrals in Eq.~\eqref{eq:KMOC_kernel_real}, and Fourier transform (c.f.~Eq.~(\ref{eq:KMOC_DObs})) to impact parameter space.
%

\Section{Soft expansion and reverse unitarity.}
\label{sec:integrals}
%
In order to efficiently evaluate the phase-space integrals appearing in the KMOC kernel (\ref{eq:KMOC_kernel_real}) we are inspired by the enormous progress in cross-section calculations in a collider physics setting where similar \emph{real} contributions appear and are handled on equal footing to the \emph{virtual} ones via \emph{reverse unitarity} \cite{Anastasiou:2002yz}, see also e.g.\ Refs.~\cite{Anastasiou:2002qz,Anastasiou:2003yy,Anastasiou:2015yha}. In the reverse unitarity setup, one replaces on-shell delta functions (and their $n$-th derivatives in intermediate steps of our calculations) that appear in phase-space integrals by the difference of propagators with varying $\imath\varepsilon$ prescription 
\begin{equation}
   \frac{2\pi \imath} {(-1)^{n}\, n!} \delta^{(n)} (z) = \frac{1}{(z-\imath\varepsilon)^{n+1}} - \frac{1}{(z+\imath\varepsilon)^{n+1}} \, , \label{eq:deltaToProp}
\end{equation}
which allows us to employ standard tools for \emph{loop integrals} like dimensional regularization, integration-by-parts (IBP) identities \cite{Chetyrkin:1981qh}, and (canonical) differential equations \cite{Kotikov:1990kg,Bern:1992em,Gehrmann:1999as,Henn:2013pwa,Henn:2014qga} to evaluate a minimal set of \emph{master integrals}. For all practical purposes, we can treat any on-shell delta function as a propagator which significantly simplifies our computations and circumvents the difficulties in having to evaluate integrals containing derivatives of delta functions that would otherwise appear.

In fact, we calculate \emph{soft integrals}, obtained by expanding the original integrals in Fig.~(\ref{fig:all_diags}) in the limit where the gravitons are much softer than the matter lines (due to the $\hbar$ scaling of momenta assigned below Eq.~(\ref{eq:kmoc_start})). Inverse graviton propagators, $\ell_i^2$, are unchanged whereas inverse matter propagators, $(\ell_i{+}p_j)^2{-}m^2$, are expanded into linearized expressions, $2 m_i u_j\cdot \ell_i$, where $u_j$ is a normalized four-velocity in the direction of the large components of the matter momentum, $p_j$. Within reverse unitarity, the phase-space delta functions are treated on the same footing and are likewise expanded. The resulting soft integrals are homogeneous in the masses and momentum transfer with the scale given by dimensional analysis, which we commonly strip from our expressions.

The construction of the differential equations for the soft master integrals $\vec{\mathcal{I}}$ has been discussed in Ref. \cite{Parra-Martinez:2020dzs}. These take the form\footnote{The RHS of the differential equation would be proportional to $\epsilon$ if we normalized all master integrals with appropriate powers of $\epsilon$, bringing it to canonical form \cite{Henn:2013pwa}. The same $\epsilon$-factorization would also apply to Eq.~(\ref{eq:diff_eq_abstract}).}
\begin{equation}
\label{eq:diff_eq_abstract}
	\mathrm{d}\vec{\mathcal{I}}(y)= A(y)\,\vec{\mathcal{I}}(y)\,,
\end{equation}
where $A(y)$ is a matrix with logarithmic singularities, $y = \sigma +{ \cal O}(q^2)$ {}\footnote{The $\mathcal O(q^2)$ shift relating $y$ and $\sigma$ is a technicality of the soft expansion and detailed in Ref.~\cite{Parra-Martinez:2020dzs}.} and $\sigma = p_1\cdot p_2/(m_1m_2)$ is the relativistic Lorentz factor. Since IBP relations are agnostic to the $\imath \varepsilon$ prescription, the connection matrix $A$ is identical for cut and virtual integrals. This allows to directly import the canonical basis constructed in Ref.~\cite{Parra-Martinez:2020dzs}.

\begin{figure}[t]
\vspace{-12pt}
	\centering
        \begin{minipage}[t]{0.32\linewidth}
     	\centering
	    \includegraphics[scale=0.75]{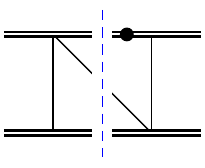}
        \end{minipage}
		\begin{minipage}[t]{0.32\linewidth}
     	\centering
	    \includegraphics[scale=0.75]{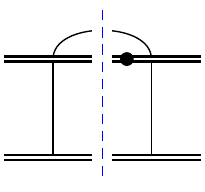}
        \end{minipage}
		\begin{minipage}[t]{0.32\linewidth}
     	\centering
	    \includegraphics[scale=0.75]{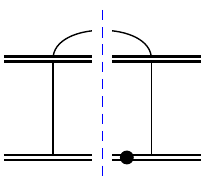}
	    \end{minipage}
	    \begin{minipage}[t]{0.32\linewidth}
     	\centering
	    \raisebox{0.35cm}{\includegraphics[scale=0.75]{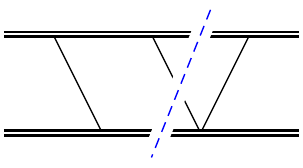}}
        \end{minipage}
		\begin{minipage}[t]{0.32\linewidth}
     	\centering
	    \raisebox{0.35cm}{\includegraphics[scale=0.75]{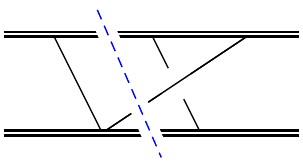}}
        \end{minipage}
		\begin{minipage}[t]{0.32\linewidth}
     	\centering
	    \raisebox{0pt}{\includegraphics[scale=0.75]{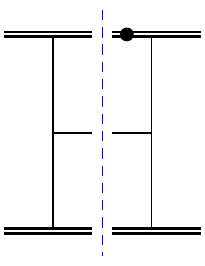}}
        \end{minipage}
        \vspace{-6pt}
\caption{Master integrals relevant for the radiated momentum. The dashed line indicates the cut; double lines, cut or uncut, are linearized propagators, and a dot ($\bullet$) indicates a squared propagator, corresponding to the $n{=}2$ case of Eq.~\eqref{eq:deltaToProp}.}
\label{fig:masters}
\end{figure}

The complete list of master integrals that survive on the triple cut relevant for the radiated momentum is illustrated in Fig.~\ref{fig:masters}. The cubic ladder-diagram with four matter propagators is notably missing from the master integrals, but will be present in the calculation of the radiation reaction on the matter lines \cite{Herrmann:2021tct}. The definition of cut integrals involving raised propagator powers is well established (see e.g.\ Ref.~\cite{Sogaard:2014ila}); one practical definition is that integration-by-parts reduction can always reduce such integrals to integrals without raised propagator powers where the meaning of cuts is clear.

All but the fourth master integral have a single s-channel Cutkosky cut (the triple cut shown in the Fig.~\ref{fig:masters}) so that the phase-space integral is simply equal to twice the imaginary part of the virtual integrals (uncut); obtained in Refs.~\cite{Herrmann:2021tct, DiVecchia:2021bdo} by solving differential equations found in Ref.~\cite{Parra-Martinez:2020dzs} but with boundary conditions evaluated in the soft rather than the potential region. The fourth cut integral in Fig.~\ref{fig:masters}, however, has to be calculated directly using reverse unitarity and differential equations. The RHS of the corresponding differential equation is proportional to the first integral in Fig.~\ref{fig:masters} \cite{Parra-Martinez:2020dzs,Herrmann:2021tct}, which we already determined above. It remains to impose a suitable boundary condition, which follows from the vanishing of such integral in the static limit $x=1$, where the real-emission phase-space volume is suppressed \cite{Herrmann:2021tct}. Then the differential equation is easily solved by direct integration with the prescribed boundary condition.
\newline\newline

%
\Section{Radiated momentum and energy loss.}
\label{sec:result}
%
Using the tools described above we have computed the radiated momentum at $\mathcal{O}(G^3)$ from the real kernel $\mathcal{I}^\mu_{R,r}$ in Eq.~(\ref{eq:KMOC_kernel_real}) and the subsequent Fourier transform to impact parameter space outlined in Eq.~(\ref{eq:KMOC_DObs}) with the following result
\begin{equation}
\label{eq:rad_mom_gen}
    \Delta R^\mu = \frac{G^3 m_1^2m_2^2}{|b|^3} \frac{u_1^\mu + u_2^\mu}{\sigma+1}   \mathcal{E}(\sigma) + {\cal O}(G^4)\,,
\end{equation}
where $u^\mu_i = p^\mu_i/m_i$, we restrict to $D=4$, and define
\begin{align}
\label{eq:rad_energy}
    \frac{\mathcal{E}(\sigma)}{\pi} =
    f_1 
    {+} f_2 \, \log \left(\frac{\sigma{+}1}{2}\right) 
    {+} f_3\,\frac{\sigma\,  \text{arcsinh}\sqrt{\frac{\sigma{-}1}{2}}}{\sqrt{\sigma^2{-}1}}\,,
\end{align}
\begin{align}
& f_1 {=}\frac{210 \sigma^6\!\!-\!552 \sigma ^5\!\!+\!339 \sigma ^4\!\!-\!912 \sigma^3\!\!+\!3148 \sigma^2\!\!-\!3336 \sigma \!+\!1151}
{48 \left(\sigma^2-1\right)^{3/2}}\,, 
\nonumber \\
&f_2{=}-\frac{35 \sigma ^4\!+60 \sigma ^3\!-150 \sigma ^2\!+76 \sigma \!-5}{8 \sqrt{\sigma ^2-1}}\,, \label{eq:fsGR}
\\
&f_3 {=}\frac{\left(2 \sigma ^2\!-3\right) \left(35 \sigma ^4 \!-30 \sigma ^2\!+11\right)}{8 (\sigma^2-1)^{3/2}}\,. \nonumber
\end{align}
Eq.~(\ref{eq:rad_mom_gen}) has the expected homogeneous mass dependence, which, as pointed out in Refs.~\cite{Kovacs:1978eu,Bini:2020hmy}, implies that the result is fixed by the probe limit $m_1\ll m_2$. Note that the result in Eq.~(\ref{eq:rad_mom_gen}) is purely longitudinal and yields the energy radiated as gravitational waves. In the center-of-mass frame it is given by
\begin{align}
\label{eq:Delta_E}
\begin{split}
    \Delta E^{\rm hyp} &{=} \frac{(p_1{+}p_2)\cdot \Delta R }{|p_1{+}p_2|} {=} \frac{G^3 M^4\nu^2}{|b|^3\, h(\nu,\sigma) }  \mathcal{E}(\sigma) {+} \mathcal{O}(G^4) \,.
\end{split}    
\end{align}
We define $h(\nu,\sigma) \equiv \sqrt{1+2\nu(\sigma-1)}$, the symmetric mass ratio $\nu \equiv m_1 m_2/M^2$, and the total mass $M\equiv(m_1{+}m_2)$. The contribution to the above energy loss from each black hole is inversely proportional to its mass. From the scattering (hyperbolic motion) result in the c.m.~frame of Eq.~(\ref{eq:Delta_E}), one can obtain the energy loss for an elliptic orbit via analytic continuation \cite{Kalin:2019rwq,Kalin:2019inp,Bini:2020hmy}
\begin{align}
\begin{split}
\label{eq:rad_energy_elliptic}
    \Delta E^{\rm ell}(\sigma, J) &= \Delta E^{\rm hyp}(\sigma,J) - \Delta E^{\rm hyp}(\sigma,-J)\\
\end{split}\,,
\end{align}
which requires writing the energy loss in terms of the angular momentum $J = b\, M\nu \sqrt{\sigma^2-1}/h(\sigma,\nu)$ and analytically continuing the result from the physical region $\sigma>1$ to the Euclidean region $\sigma<1$, where $\sigma$ is related to the dimensionless binding energy $\overline{E}{=}\frac{h(\nu,\sigma){-}1}{\nu}{<}0$ \cite{Bini:2020hmy}
\begin{equation}
\label{eq:ell_energy_loss_more_explicit}
    \Delta E^{\rm ell}(\sigma, J) = \frac{G^3 M^7 \nu^5 (1-\sigma^2)^{\frac{3}{2}}}{J^3\, h(\nu,\sigma)^4} \, \widetilde{\mathcal{E}}^{\rm ell}(\sigma) + \mathcal{O}(G^4)\,.
\end{equation}
We define an analogous rescaled function $\widetilde{\mathcal{E}}^{\rm ell}(\sigma)$
\begin{align}
    \frac{\widetilde{\mathcal{E}}^{\rm ell}(\sigma)}{\pi} = 
      \widetilde{f}_1 
    {-} \widetilde{f}_2 \, \log \left(\frac{\sigma{+}1}{2}\right) 
    {+} \widetilde{f}_3\,\frac{\sigma\,   \text{arcsin}\sqrt{\frac{1-\sigma}{2}}}{\sqrt{1{-}\sigma^2}}\,, 
\end{align}
with $\widetilde{f}_i {=} 2 f_i$, and $f_i$ given in Eq.~\eqref{eq:fsGR} subject to the additional replacement $(\sigma^2\!{-}1)^{\frac{n}{2}} {\rightarrow} (1{-}\sigma^2)^{\frac{n}{2}}$ for odd integers~$n$. Note that the elliptic orbit energy loss presented in Eq.~(\ref{eq:ell_energy_loss_more_explicit}) has the expected simplified $\nu$ dependence observed by \cite{Bini:2020hmy} that is inherited from the analytic continuation of the hyperbolic result.

\Section{Cross-checks.}
%
Our result for the energy loss for scattering black holes can be expanded in small velocity ~$v = \sqrt{\sigma^2-1}/\sigma$
\begin{equation}
\label{eq:rad_e_static_exp}
\hspace{-.1cm}
    \!\frac{\mathcal{E}(\sigma)}{\pi} {=}  
    \frac{37}{15} v{+}\frac{2393}{840} v^3\!{+}\frac{61703}{10080} v^5
    \!{+}\frac{3131839}{354816} v^7
    \!{+}\mathcal{O}(v^{9}),
\hspace{-.4cm}    
\end{equation}
and compared to known post-Newtonian (PN) data.
The first three terms in Eq.~\eqref{eq:rad_e_static_exp} are found to agree with the result known up to 2PN \cite{Kovacs:1978eu, Blanchet:1989cu,Bini:2020hmy}. We can also compare the energy loss for elliptic orbits in Eq.~(\ref{eq:ell_energy_loss_more_explicit}) for small velocities to the 3PN accurate results for the instantaneous energy flux integrated over an orbit from Refs.~\cite{Peters:1963ux,Peters:1964zz,Wagoner:1976am,Blanchet:1989cu,Junker:1992kle,Gopakumar:1997ng,Gopakumar:2001dy,Arun:2007sg,Blanchet:2013haa} in the large eccentricity limit, i.e.~to leading order in large $J$. The velocity expansion is equivalent to Eq.~\eqref{eq:rad_e_static_exp} (up to a factor of 2) since it is controlled by the same analytic function and we find perfect agreement where our results overlap. Note that the full 3PN flux includes tail (or hereditary) contributions at 1.5, 2.5 and 3PN order which are known analytically only in the limit of small eccentricity \cite{Blanchet:1993ec,Rieth:1997mk,Arun:2007rg}. The agreement of our result with the instantaneous part suggests that the tail contributions must be sub-leading at large eccentricity.

Going back to the hyperbolic orbit, instead of the low-velocity result, we can also compare the ultra-relativistic limit $\sigma \rightarrow \infty$ of Eq.~(\ref{eq:rad_energy}). The apparent logarithmic divergence cancels and one finds
\begin{equation}
    \mathcal{E}(\sigma) =  \frac{35}{8} \pi  (1+2 \log2) \sigma^3 + \mathcal{O} \left(\sigma^2\right)\,.
    \label{eq:EheGR}
\end{equation}
This can be compared to the prediction by Kovacs and Thorne \cite{Kovacs:1978eu}, based upon the numerical probe calculation by Peters \cite{Peters:1970mx}. Both expressions agree structurally, but disagree in the numerical coefficient\,\footnote{After our work appeared on the arXiv, we were informed by T.~Damour of his numerical computation, together with Bini and Geralico, of the high-energy coefficient, which is in agreement with our analytic result. Furthermore, they extended their computation in the small velocity limit up to $\mathcal{O}(v^{15})$, also finding agreement with our results. See also Refs.~\cite{Jakobsen:2021smu, Mougiakakos:2021ckm} for a confirmation of our results up to $\mathcal{O}(v^{7})$ using worldline techniques.}. We note that our results are only valid for $\sigma \ll (GE_{\rm cm}/b)^{-1}$, beyond which perturbation theory breaks down. This is evidentiated by the fact that using Eq.~\eqref{eq:EheGR} one would conclude that at large enough $\sigma$ the radiated energy exceeds the incoming energy, which is nonsense. This can be interpreted as signaling the necessity of accounting for destructive interference in multi-graviton emission, which cuts off the spectrum of gravitational waves at high-frequency\,\footnote{We thank Gabriele Veneziano for discussions on this point.}, as explained in Refs.~\cite{Gruzinov:2014moa,Ciafaloni:2015xsr,Ciafaloni:2018uwe}.

In addition, we have compared our result in Eq.~(\ref{eq:Delta_E}) with the coefficient of the tail term in the ${\cal O}(G^4)$ radial action of Ref.~\cite{Bern:2021dqo}, which is proportional to $\Delta E$ \cite{Bini:2017wfr,Blanchet:2019rjs,Bini:2020hmy}, finding full agreement.

\Section{(Non)-Universality.}
%
At ${\cal O}(G^3)$ it has been shown that the gravitational deflection angle has universal properties in the ultra-relativistic limit \cite{Amati:1990xe,Bern:2020gjj,DiVecchia:2020ymx,Damour:2020tta}. We have also computed the radiated momentum in ${\cal N} = 8$ supergravity \cite{Cremmer:1979up} (for BPS angle $\phi=\pi/2$ \cite{Caron-Huot:2018ape}), using the integrand in Ref.~\cite{Parra-Martinez:2020dzs} and the technology described above. The result has the same structure as Eqs.~\eqref{eq:rad_energy}--\eqref{eq:fsGR} with
\begin{align}
\begin{split}
     \hspace{-.2cm}
      f_1 {=} \frac{8\sigma^6}{(\sigma^2{-}1)^{3/2}}\,,\, 
      f_2 {=}-\frac{8\sigma^4}{\sqrt{\sigma^2{-}1}} \,, \, 
      f_3 {=} \frac{16\sigma^4(\sigma^2{-}2)}{(\sigma^2{-}1)^{3/2}}\,. 
      \hspace{-.1cm}
\end{split}
\end{align}
As in pure gravity, the ultra-relativistic limit of the radiated momentum is controlled by the combinations $f_1$ and $-f_2+f_3/2$ (with the leading term independent of $\phi$). Although the limit does not coincide with Eq.~\eqref{eq:EheGR} in its rational prefactor (35/8 vs. 8), we note that the ratio of the logarithmic ($\log 2$) and non-logarithmic contributions appears universal, i.e.,
\begin{equation}
  \lim_{\sigma \rightarrow \infty} \frac{-2f_2+f_3}{2\,f_1}\Big|_{{\cal N} = 8} =  \lim_{\sigma \rightarrow \infty} \frac{-2f_2+f_3}{2\, f_1}\Big|_{{\rm GR}} = 2\,.
\end{equation}

\Section{Conclusions.}
%
In this letter we report our computation of the radiated energy emitted in gravitational waves during the scattering of two spinless black holes in general relativity to leading order in Newton's constant and all orders in velocity. Furthermore, we obtain the radiated energy in elliptic orbits by analytic continuation from the scattering problem \cite{Bini:2020hmy}.  Expanding our results in small velocity we find perfect agreement with the known PN data \cite{Kovacs:1978eu,Peters:1970mx}. In the high-energy regime, we agree with the kinematic dependence described in Ref.~\cite{Kovacs:1978eu} but disagree with their numerical coefficient.

Besides the radiated momentum discussed here, the KMOC formalism can also be used to calculate the transverse impulse on individual particles, which yields the deflection angle \cite{Bern:2019crd, Bern:2019nnu, Kalin:2020mvi, Kalin:2020fhe}, including radiation reaction \cite{DiVecchia:2020ymx,Damour:2020tta,DiVecchia:2021ndb}. This computation is more involved than the one presented here, because it requires the full virtual soft amplitude, so we defer its discussion. 
The tools described here can be directly used to compute observables for spinning \cite{Maybee:2019jus}, and charged \cite{delaCruz:2020bbn} black holes.  Furthermore, in retaining the collider-physics analogy, and by restricting the integration over phase-space, one can imagine computing differential observables, such as the radiated energy spectrum, or the angular energy distribution \cite{Gonzo:2020xza}, analogous to e.g.~rapidity distributions (see~Ref.~\cite{Anastasiou:2002qz}) or energy correlators (see~Ref.~\cite{Korchemsky:1997sy, Korchemsky:1999kt}).  We leave the discussion of such observables to future work.

\Section{Acknowledgments:} 
%
We thank Zvi Bern, Radu Roiban, Chia-Hsien Shen, and Mikhail Solon for helpful comments and collaboration on related projects, and Clifford Cheung and Rafael Porto for discussions and comments on the draft. 
We are also grateful to Paolo Di Vecchia,  Carlo Heissenberg, Rodolfo Russo, and Gabriele Veneziano for stimulating discussions regarding radiative effects and cross-checks of various soft master integrals. 
E.H. thanks Lance Dixon and Bernhard Mistlberger for discussions about the reverse unitarity method. 
E.H.\ is supported by the U.S.\ Department of Energy (DOE) under Award Number DE-SC0009937. 
J.P.-M.\ is supported by the U.S.\ Department of Energy (DOE) under Award Number~DE-SC0011632. 
M.S.R.’s work is funded by the German Research Foundation (DFG) within the Research Training Group GRK 2044. 
M.Z.'s work is funded by the U.K.\ Royal Society through Grant URF{\textbackslash}R1{\textbackslash}20109.


\vspace{-0.5cm}
\bibliographystyle{apsrev4-1} 
\bibliography{erad_refs_v2} 

\begin{thebibliography}{126}%
\makeatletter
\providecommand \@ifxundefined [1]{%
 \@ifx{#1\undefined}
}%
\providecommand \@ifnum [1]{%
 \ifnum #1\expandafter \@firstoftwo
 \else \expandafter \@secondoftwo
 \fi
}%
\providecommand \@ifx [1]{%
 \ifx #1\expandafter \@firstoftwo
 \else \expandafter \@secondoftwo
 \fi
}%
\providecommand \natexlab [1]{#1}%
\providecommand \enquote  [1]{``#1''}%
\providecommand \bibnamefont  [1]{#1}%
\providecommand \bibfnamefont [1]{#1}%
\providecommand \citenamefont [1]{#1}%
\providecommand \href@noop [0]{\@secondoftwo}%
\providecommand \href [0]{\begingroup \@sanitize@url \@href}%
\providecommand \@href[1]{\@@startlink{#1}\@@href}%
\providecommand \@@href[1]{\endgroup#1\@@endlink}%
\providecommand \@sanitize@url [0]{\catcode `\\12\catcode `\$12\catcode
  `\&12\catcode `\#12\catcode `\^12\catcode `\_12\catcode `\%12\relax}%
\providecommand \@@startlink[1]{}%
\providecommand \@@endlink[0]{}%
\providecommand \url  [0]{\begingroup\@sanitize@url \@url }%
\providecommand \@url [1]{\endgroup\@href {#1}{\urlprefix }}%
\providecommand \urlprefix  [0]{URL }%
\providecommand \Eprint [0]{\href }%
\providecommand \doibase [0]{http://dx.doi.org/}%
\providecommand \selectlanguage [0]{\@gobble}%
\providecommand \bibinfo  [0]{\@secondoftwo}%
\providecommand \bibfield  [0]{\@secondoftwo}%
\providecommand \translation [1]{[#1]}%
\providecommand \BibitemOpen [0]{}%
\providecommand \bibitemStop [0]{}%
\providecommand \bibitemNoStop [0]{.\EOS\space}%
\providecommand \EOS [0]{\spacefactor3000\relax}%
\providecommand \BibitemShut  [1]{\csname bibitem#1\endcsname}%
\let\auto@bib@innerbib\@empty
\bibitem [{\citenamefont {Bern}\ \emph {et~al.}(1994)\citenamefont {Bern},
  \citenamefont {Dixon}, \citenamefont {Dunbar},\ and\ \citenamefont
  {Kosower}}]{Bern:1994zx}%
  \BibitemOpen
  \bibfield  {author} {\bibinfo {author} {\bibfnamefont {Z.}~\bibnamefont
  {Bern}}, \bibinfo {author} {\bibfnamefont {L.~J.}\ \bibnamefont {Dixon}},
  \bibinfo {author} {\bibfnamefont {D.~C.}\ \bibnamefont {Dunbar}}, \ and\
  \bibinfo {author} {\bibfnamefont {D.~A.}\ \bibnamefont {Kosower}},\ }\href
  {\doibase 10.1016/0550-3213(94)90179-1} {\bibfield  {journal} {\bibinfo
  {journal} {Nucl. Phys. B}\ }\textbf {\bibinfo {volume} {425}},\ \bibinfo
  {pages} {217} (\bibinfo {year} {1994})},\ \Eprint
  {http://arxiv.org/abs/hep-ph/9403226} {arXiv:hep-ph/9403226} \BibitemShut
  {NoStop}%
\bibitem [{\citenamefont {Bern}\ \emph {et~al.}(1995)\citenamefont {Bern},
  \citenamefont {Dixon}, \citenamefont {Dunbar},\ and\ \citenamefont
  {Kosower}}]{Bern:1994cg}%
  \BibitemOpen
  \bibfield  {author} {\bibinfo {author} {\bibfnamefont {Z.}~\bibnamefont
  {Bern}}, \bibinfo {author} {\bibfnamefont {L.~J.}\ \bibnamefont {Dixon}},
  \bibinfo {author} {\bibfnamefont {D.~C.}\ \bibnamefont {Dunbar}}, \ and\
  \bibinfo {author} {\bibfnamefont {D.~A.}\ \bibnamefont {Kosower}},\ }\href
  {\doibase 10.1016/0550-3213(94)00488-Z} {\bibfield  {journal} {\bibinfo
  {journal} {Nucl. Phys. B}\ }\textbf {\bibinfo {volume} {435}},\ \bibinfo
  {pages} {59} (\bibinfo {year} {1995})},\ \Eprint
  {http://arxiv.org/abs/hep-ph/9409265} {arXiv:hep-ph/9409265} \BibitemShut
  {NoStop}%
\bibitem [{\citenamefont {Britto}\ \emph {et~al.}(2005)\citenamefont {Britto},
  \citenamefont {Cachazo},\ and\ \citenamefont {Feng}}]{Britto:2004nc}%
  \BibitemOpen
  \bibfield  {author} {\bibinfo {author} {\bibfnamefont {R.}~\bibnamefont
  {Britto}}, \bibinfo {author} {\bibfnamefont {F.}~\bibnamefont {Cachazo}}, \
  and\ \bibinfo {author} {\bibfnamefont {B.}~\bibnamefont {Feng}},\ }\href
  {\doibase 10.1016/j.nuclphysb.2005.07.014} {\bibfield  {journal} {\bibinfo
  {journal} {Nucl. Phys. B}\ }\textbf {\bibinfo {volume} {725}},\ \bibinfo
  {pages} {275} (\bibinfo {year} {2005})},\ \Eprint
  {http://arxiv.org/abs/hep-th/0412103} {arXiv:hep-th/0412103} \BibitemShut
  {NoStop}%
\bibitem [{\citenamefont {Bern}\ \emph {et~al.}(2008)\citenamefont {Bern},
  \citenamefont {Carrasco},\ and\ \citenamefont {Johansson}}]{Bern:2008qj}%
  \BibitemOpen
  \bibfield  {author} {\bibinfo {author} {\bibfnamefont {Z.}~\bibnamefont
  {Bern}}, \bibinfo {author} {\bibfnamefont {J.}~\bibnamefont {Carrasco}}, \
  and\ \bibinfo {author} {\bibfnamefont {H.}~\bibnamefont {Johansson}},\ }\href
  {\doibase 10.1103/PhysRevD.78.085011} {\bibfield  {journal} {\bibinfo
  {journal} {Phys. Rev. D}\ }\textbf {\bibinfo {volume} {78}},\ \bibinfo
  {pages} {085011} (\bibinfo {year} {2008})},\ \Eprint
  {http://arxiv.org/abs/0805.3993} {arXiv:0805.3993 [hep-ph]} \BibitemShut
  {NoStop}%
\bibitem [{\citenamefont {Bern}\ \emph {et~al.}(2010)\citenamefont {Bern},
  \citenamefont {Carrasco},\ and\ \citenamefont {Johansson}}]{Bern:2010ue}%
  \BibitemOpen
  \bibfield  {author} {\bibinfo {author} {\bibfnamefont {Z.}~\bibnamefont
  {Bern}}, \bibinfo {author} {\bibfnamefont {J.~J.~M.}\ \bibnamefont
  {Carrasco}}, \ and\ \bibinfo {author} {\bibfnamefont {H.}~\bibnamefont
  {Johansson}},\ }\href {\doibase 10.1103/PhysRevLett.105.061602} {\bibfield
  {journal} {\bibinfo  {journal} {Phys. Rev. Lett.}\ }\textbf {\bibinfo
  {volume} {105}},\ \bibinfo {pages} {061602} (\bibinfo {year} {2010})},\
  \Eprint {http://arxiv.org/abs/1004.0476} {arXiv:1004.0476 [hep-th]}
  \BibitemShut {NoStop}%
\bibitem [{\citenamefont {Bern}\ \emph {et~al.}(2012)\citenamefont {Bern},
  \citenamefont {Carrasco}, \citenamefont {Dixon}, \citenamefont {Johansson},\
  and\ \citenamefont {Roiban}}]{Bern:2012uf}%
  \BibitemOpen
  \bibfield  {author} {\bibinfo {author} {\bibfnamefont {Z.}~\bibnamefont
  {Bern}}, \bibinfo {author} {\bibfnamefont {J.}~\bibnamefont {Carrasco}},
  \bibinfo {author} {\bibfnamefont {L.}~\bibnamefont {Dixon}}, \bibinfo
  {author} {\bibfnamefont {H.}~\bibnamefont {Johansson}}, \ and\ \bibinfo
  {author} {\bibfnamefont {R.}~\bibnamefont {Roiban}},\ }\href {\doibase
  10.1103/PhysRevD.85.105014} {\bibfield  {journal} {\bibinfo  {journal} {Phys.
  Rev. D}\ }\textbf {\bibinfo {volume} {85}},\ \bibinfo {pages} {105014}
  (\bibinfo {year} {2012})},\ \Eprint {http://arxiv.org/abs/1201.5366}
  {arXiv:1201.5366 [hep-th]} \BibitemShut {NoStop}%
\bibitem [{\citenamefont {Bern}\ \emph {et~al.}(2017)\citenamefont {Bern},
  \citenamefont {Carrasco}, \citenamefont {Chen}, \citenamefont {Johansson},
  \citenamefont {Roiban},\ and\ \citenamefont {Zeng}}]{Bern:2017ucb}%
  \BibitemOpen
  \bibfield  {author} {\bibinfo {author} {\bibfnamefont {Z.}~\bibnamefont
  {Bern}}, \bibinfo {author} {\bibfnamefont {J.~J.~M.}\ \bibnamefont
  {Carrasco}}, \bibinfo {author} {\bibfnamefont {W.-M.}\ \bibnamefont {Chen}},
  \bibinfo {author} {\bibfnamefont {H.}~\bibnamefont {Johansson}}, \bibinfo
  {author} {\bibfnamefont {R.}~\bibnamefont {Roiban}}, \ and\ \bibinfo {author}
  {\bibfnamefont {M.}~\bibnamefont {Zeng}},\ }\href {\doibase
  10.1103/PhysRevD.96.126012} {\bibfield  {journal} {\bibinfo  {journal} {Phys.
  Rev. D}\ }\textbf {\bibinfo {volume} {96}},\ \bibinfo {pages} {126012}
  (\bibinfo {year} {2017})},\ \Eprint {http://arxiv.org/abs/1708.06807}
  {arXiv:1708.06807 [hep-th]} \BibitemShut {NoStop}%
\bibitem [{\citenamefont {Bern}\ \emph {et~al.}(2018)\citenamefont {Bern},
  \citenamefont {Carrasco}, \citenamefont {Chen}, \citenamefont {Edison},
  \citenamefont {Johansson}, \citenamefont {Parra-Martinez}, \citenamefont
  {Roiban},\ and\ \citenamefont {Zeng}}]{Bern:2018jmv}%
  \BibitemOpen
  \bibfield  {author} {\bibinfo {author} {\bibfnamefont {Z.}~\bibnamefont
  {Bern}}, \bibinfo {author} {\bibfnamefont {J.~J.}\ \bibnamefont {Carrasco}},
  \bibinfo {author} {\bibfnamefont {W.-M.}\ \bibnamefont {Chen}}, \bibinfo
  {author} {\bibfnamefont {A.}~\bibnamefont {Edison}}, \bibinfo {author}
  {\bibfnamefont {H.}~\bibnamefont {Johansson}}, \bibinfo {author}
  {\bibfnamefont {J.}~\bibnamefont {Parra-Martinez}}, \bibinfo {author}
  {\bibfnamefont {R.}~\bibnamefont {Roiban}}, \ and\ \bibinfo {author}
  {\bibfnamefont {M.}~\bibnamefont {Zeng}},\ }\href {\doibase
  10.1103/PhysRevD.98.086021} {\bibfield  {journal} {\bibinfo  {journal} {Phys.
  Rev. D}\ }\textbf {\bibinfo {volume} {98}},\ \bibinfo {pages} {086021}
  (\bibinfo {year} {2018})},\ \Eprint {http://arxiv.org/abs/1804.09311}
  {arXiv:1804.09311 [hep-th]} \BibitemShut {NoStop}%
\bibitem [{\citenamefont {Bern}\ \emph
  {et~al.}(2019{\natexlab{a}})\citenamefont {Bern}, \citenamefont {Carrasco},
  \citenamefont {Chiodaroli}, \citenamefont {Johansson},\ and\ \citenamefont
  {Roiban}}]{Bern:2019prr}%
  \BibitemOpen
  \bibfield  {author} {\bibinfo {author} {\bibfnamefont {Z.}~\bibnamefont
  {Bern}}, \bibinfo {author} {\bibfnamefont {J.~J.}\ \bibnamefont {Carrasco}},
  \bibinfo {author} {\bibfnamefont {M.}~\bibnamefont {Chiodaroli}}, \bibinfo
  {author} {\bibfnamefont {H.}~\bibnamefont {Johansson}}, \ and\ \bibinfo
  {author} {\bibfnamefont {R.}~\bibnamefont {Roiban}},\ }\href@noop {} {\
  (\bibinfo {year} {2019}{\natexlab{a}})},\ \Eprint
  {http://arxiv.org/abs/1909.01358} {arXiv:1909.01358 [hep-th]} \BibitemShut
  {NoStop}%
\bibitem [{\citenamefont {Goldberger}\ and\ \citenamefont
  {Rothstein}(2006)}]{Goldberger:2004jt}%
  \BibitemOpen
  \bibfield  {author} {\bibinfo {author} {\bibfnamefont {W.~D.}\ \bibnamefont
  {Goldberger}}\ and\ \bibinfo {author} {\bibfnamefont {I.~Z.}\ \bibnamefont
  {Rothstein}},\ }\href {\doibase 10.1103/PhysRevD.73.104029} {\bibfield
  {journal} {\bibinfo  {journal} {Phys. Rev. D}\ }\textbf {\bibinfo {volume}
  {73}},\ \bibinfo {pages} {104029} (\bibinfo {year} {2006})},\ \Eprint
  {http://arxiv.org/abs/hep-th/0409156} {arXiv:hep-th/0409156} \BibitemShut
  {NoStop}%
\bibitem [{\citenamefont {Porto}(2016)}]{Porto:2016pyg}%
  \BibitemOpen
  \bibfield  {author} {\bibinfo {author} {\bibfnamefont {R.~A.}\ \bibnamefont
  {Porto}},\ }\href {\doibase 10.1016/j.physrep.2016.04.003} {\bibfield
  {journal} {\bibinfo  {journal} {Phys. Rept.}\ }\textbf {\bibinfo {volume}
  {633}},\ \bibinfo {pages} {1} (\bibinfo {year} {2016})},\ \Eprint
  {http://arxiv.org/abs/1601.04914} {arXiv:1601.04914 [hep-th]} \BibitemShut
  {NoStop}%
\bibitem [{\citenamefont {Cheung}\ \emph {et~al.}(2018)\citenamefont {Cheung},
  \citenamefont {Rothstein},\ and\ \citenamefont {Solon}}]{Cheung:2018wkq}%
  \BibitemOpen
  \bibfield  {author} {\bibinfo {author} {\bibfnamefont {C.}~\bibnamefont
  {Cheung}}, \bibinfo {author} {\bibfnamefont {I.~Z.}\ \bibnamefont
  {Rothstein}}, \ and\ \bibinfo {author} {\bibfnamefont {M.~P.}\ \bibnamefont
  {Solon}},\ }\href {\doibase 10.1103/PhysRevLett.121.251101} {\bibfield
  {journal} {\bibinfo  {journal} {Phys. Rev. Lett.}\ }\textbf {\bibinfo
  {volume} {121}},\ \bibinfo {pages} {251101} (\bibinfo {year} {2018})},\
  \Eprint {http://arxiv.org/abs/1808.02489} {arXiv:1808.02489 [hep-th]}
  \BibitemShut {NoStop}%
\bibitem [{\citenamefont {Abbott}\ \emph {et~al.}(2016)\citenamefont {Abbott}
  \emph {et~al.}}]{Abbott:2016blz}%
  \BibitemOpen
  \bibfield  {author} {\bibinfo {author} {\bibfnamefont {B.}~\bibnamefont
  {Abbott}} \emph {et~al.} (\bibinfo {collaboration} {LIGO Scientific,
  Virgo}),\ }\href {\doibase 10.1103/PhysRevLett.116.061102} {\bibfield
  {journal} {\bibinfo  {journal} {Phys. Rev. Lett.}\ }\textbf {\bibinfo
  {volume} {116}},\ \bibinfo {pages} {061102} (\bibinfo {year} {2016})},\
  \Eprint {http://arxiv.org/abs/1602.03837} {arXiv:1602.03837 [gr-qc]}
  \BibitemShut {NoStop}%
\bibitem [{\citenamefont {Abbott}\ \emph {et~al.}(2017)\citenamefont {Abbott}
  \emph {et~al.}}]{TheLIGOScientific:2017qsa}%
  \BibitemOpen
  \bibfield  {author} {\bibinfo {author} {\bibfnamefont {B.}~\bibnamefont
  {Abbott}} \emph {et~al.} (\bibinfo {collaboration} {LIGO Scientific,
  Virgo}),\ }\href {\doibase 10.1103/PhysRevLett.119.161101} {\bibfield
  {journal} {\bibinfo  {journal} {Phys. Rev. Lett.}\ }\textbf {\bibinfo
  {volume} {119}},\ \bibinfo {pages} {161101} (\bibinfo {year} {2017})},\
  \Eprint {http://arxiv.org/abs/1710.05832} {arXiv:1710.05832 [gr-qc]}
  \BibitemShut {NoStop}%
\bibitem [{\citenamefont {Punturo}\ \emph {et~al.}(2010)\citenamefont {Punturo}
  \emph {et~al.}}]{Punturo_2010}%
  \BibitemOpen
  \bibfield  {author} {\bibinfo {author} {\bibfnamefont {M.}~\bibnamefont
  {Punturo}} \emph {et~al.},\ }\href {\doibase 10.1088/0264-9381/27/19/194002}
  {\bibfield  {journal} {\bibinfo  {journal} {Classical and Quantum Gravity}\
  }\textbf {\bibinfo {volume} {27}},\ \bibinfo {pages} {194002} (\bibinfo
  {year} {2010})}\BibitemShut {NoStop}%
\bibitem [{\citenamefont {Amaro-Seoane}\ \emph {et~al.}(2017)\citenamefont
  {Amaro-Seoane} \emph {et~al.}}]{Audley:2017drz}%
  \BibitemOpen
  \bibfield  {author} {\bibinfo {author} {\bibfnamefont {P.}~\bibnamefont
  {Amaro-Seoane}} \emph {et~al.} (\bibinfo {collaboration} {LISA}),\
  }\href@noop {} {\  (\bibinfo {year} {2017})},\ \Eprint
  {http://arxiv.org/abs/1702.00786} {arXiv:1702.00786 [astro-ph.IM]}
  \BibitemShut {NoStop}%
\bibitem [{\citenamefont {Reitze}\ \emph {et~al.}(2019)\citenamefont {Reitze}
  \emph {et~al.}}]{Reitze:2019iox}%
  \BibitemOpen
  \bibfield  {author} {\bibinfo {author} {\bibfnamefont {D.}~\bibnamefont
  {Reitze}} \emph {et~al.},\ }\href@noop {} {\bibfield  {journal} {\bibinfo
  {journal} {Bull. Am. Astron. Soc.}\ }\textbf {\bibinfo {volume} {51}},\
  \bibinfo {pages} {035} (\bibinfo {year} {2019})},\ \Eprint
  {http://arxiv.org/abs/1907.04833} {arXiv:1907.04833 [astro-ph.IM]}
  \BibitemShut {NoStop}%
\bibitem [{\citenamefont {Damour}(2016)}]{Damour:2016gwp}%
  \BibitemOpen
  \bibfield  {author} {\bibinfo {author} {\bibfnamefont {T.}~\bibnamefont
  {Damour}},\ }\href {\doibase 10.1103/PhysRevD.94.104015} {\bibfield
  {journal} {\bibinfo  {journal} {Phys. Rev. D}\ }\textbf {\bibinfo {volume}
  {94}},\ \bibinfo {pages} {104015} (\bibinfo {year} {2016})},\ \Eprint
  {http://arxiv.org/abs/1609.00354} {arXiv:1609.00354 [gr-qc]} \BibitemShut
  {NoStop}%
\bibitem [{\citenamefont {Foffa}\ \emph {et~al.}(2017)\citenamefont {Foffa},
  \citenamefont {Mastrolia}, \citenamefont {Sturani},\ and\ \citenamefont
  {Sturm}}]{Foffa:2016rgu}%
  \BibitemOpen
  \bibfield  {author} {\bibinfo {author} {\bibfnamefont {S.}~\bibnamefont
  {Foffa}}, \bibinfo {author} {\bibfnamefont {P.}~\bibnamefont {Mastrolia}},
  \bibinfo {author} {\bibfnamefont {R.}~\bibnamefont {Sturani}}, \ and\
  \bibinfo {author} {\bibfnamefont {C.}~\bibnamefont {Sturm}},\ }\href
  {\doibase 10.1103/PhysRevD.95.104009} {\bibfield  {journal} {\bibinfo
  {journal} {Phys. Rev. D}\ }\textbf {\bibinfo {volume} {95}},\ \bibinfo
  {pages} {104009} (\bibinfo {year} {2017})},\ \Eprint
  {http://arxiv.org/abs/1612.00482} {arXiv:1612.00482 [gr-qc]} \BibitemShut
  {NoStop}%
\bibitem [{\citenamefont {Bl\"umlein}\ \emph
  {et~al.}(2020{\natexlab{a}})\citenamefont {Bl\"umlein}, \citenamefont
  {Maier},\ and\ \citenamefont {Marquard}}]{Blumlein:2019zku}%
  \BibitemOpen
  \bibfield  {author} {\bibinfo {author} {\bibfnamefont {J.}~\bibnamefont
  {Bl\"umlein}}, \bibinfo {author} {\bibfnamefont {A.}~\bibnamefont {Maier}}, \
  and\ \bibinfo {author} {\bibfnamefont {P.}~\bibnamefont {Marquard}},\ }\href
  {\doibase 10.1016/j.physletb.2019.135100} {\bibfield  {journal} {\bibinfo
  {journal} {Phys. Lett. B}\ }\textbf {\bibinfo {volume} {800}},\ \bibinfo
  {pages} {135100} (\bibinfo {year} {2020}{\natexlab{a}})},\ \Eprint
  {http://arxiv.org/abs/1902.11180} {arXiv:1902.11180 [gr-qc]} \BibitemShut
  {NoStop}%
\bibitem [{\citenamefont {Bl\"umlein}\ \emph
  {et~al.}(2020{\natexlab{b}})\citenamefont {Bl\"umlein}, \citenamefont
  {Maier}, \citenamefont {Marquard},\ and\ \citenamefont
  {Sch\"afer}}]{Blumlein:2020znm}%
  \BibitemOpen
  \bibfield  {author} {\bibinfo {author} {\bibfnamefont {J.}~\bibnamefont
  {Bl\"umlein}}, \bibinfo {author} {\bibfnamefont {A.}~\bibnamefont {Maier}},
  \bibinfo {author} {\bibfnamefont {P.}~\bibnamefont {Marquard}}, \ and\
  \bibinfo {author} {\bibfnamefont {G.}~\bibnamefont {Sch\"afer}},\ }\href
  {\doibase 10.1016/j.physletb.2020.135496} {\bibfield  {journal} {\bibinfo
  {journal} {Phys. Lett. B}\ }\textbf {\bibinfo {volume} {807}},\ \bibinfo
  {pages} {135496} (\bibinfo {year} {2020}{\natexlab{b}})},\ \Eprint
  {http://arxiv.org/abs/2003.07145} {arXiv:2003.07145 [gr-qc]} \BibitemShut
  {NoStop}%
\bibitem [{\citenamefont {Bjerrum-Bohr}\ \emph {et~al.}(2018)\citenamefont
  {Bjerrum-Bohr}, \citenamefont {Damgaard}, \citenamefont {Festuccia},
  \citenamefont {Plant\'e},\ and\ \citenamefont
  {Vanhove}}]{Bjerrum-Bohr:2018xdl}%
  \BibitemOpen
  \bibfield  {author} {\bibinfo {author} {\bibfnamefont {N.~J.}\ \bibnamefont
  {Bjerrum-Bohr}}, \bibinfo {author} {\bibfnamefont {P.~H.}\ \bibnamefont
  {Damgaard}}, \bibinfo {author} {\bibfnamefont {G.}~\bibnamefont {Festuccia}},
  \bibinfo {author} {\bibfnamefont {L.}~\bibnamefont {Plant\'e}}, \ and\
  \bibinfo {author} {\bibfnamefont {P.}~\bibnamefont {Vanhove}},\ }\href
  {\doibase 10.1103/PhysRevLett.121.171601} {\bibfield  {journal} {\bibinfo
  {journal} {Phys. Rev. Lett.}\ }\textbf {\bibinfo {volume} {121}},\ \bibinfo
  {pages} {171601} (\bibinfo {year} {2018})},\ \Eprint
  {http://arxiv.org/abs/1806.04920} {arXiv:1806.04920 [hep-th]} \BibitemShut
  {NoStop}%
\bibitem [{\citenamefont {Bern}\ \emph
  {et~al.}(2019{\natexlab{b}})\citenamefont {Bern}, \citenamefont {Cheung},
  \citenamefont {Roiban}, \citenamefont {Shen}, \citenamefont {Solon},\ and\
  \citenamefont {Zeng}}]{Bern:2019nnu}%
  \BibitemOpen
  \bibfield  {author} {\bibinfo {author} {\bibfnamefont {Z.}~\bibnamefont
  {Bern}}, \bibinfo {author} {\bibfnamefont {C.}~\bibnamefont {Cheung}},
  \bibinfo {author} {\bibfnamefont {R.}~\bibnamefont {Roiban}}, \bibinfo
  {author} {\bibfnamefont {C.-H.}\ \bibnamefont {Shen}}, \bibinfo {author}
  {\bibfnamefont {M.~P.}\ \bibnamefont {Solon}}, \ and\ \bibinfo {author}
  {\bibfnamefont {M.}~\bibnamefont {Zeng}},\ }\href {\doibase
  10.1103/PhysRevLett.122.201603} {\bibfield  {journal} {\bibinfo  {journal}
  {Phys. Rev. Lett.}\ }\textbf {\bibinfo {volume} {122}},\ \bibinfo {pages}
  {201603} (\bibinfo {year} {2019}{\natexlab{b}})},\ \Eprint
  {http://arxiv.org/abs/1901.04424} {arXiv:1901.04424 [hep-th]} \BibitemShut
  {NoStop}%
\bibitem [{\citenamefont {Foffa}\ \emph {et~al.}(2019)\citenamefont {Foffa},
  \citenamefont {Mastrolia}, \citenamefont {Sturani}, \citenamefont {Sturm},\
  and\ \citenamefont {Torres~Bobadilla}}]{Foffa:2019hrb}%
  \BibitemOpen
  \bibfield  {author} {\bibinfo {author} {\bibfnamefont {S.}~\bibnamefont
  {Foffa}}, \bibinfo {author} {\bibfnamefont {P.}~\bibnamefont {Mastrolia}},
  \bibinfo {author} {\bibfnamefont {R.}~\bibnamefont {Sturani}}, \bibinfo
  {author} {\bibfnamefont {C.}~\bibnamefont {Sturm}}, \ and\ \bibinfo {author}
  {\bibfnamefont {W.~J.}\ \bibnamefont {Torres~Bobadilla}},\ }\href {\doibase
  10.1103/PhysRevLett.122.241605} {\bibfield  {journal} {\bibinfo  {journal}
  {Phys. Rev. Lett.}\ }\textbf {\bibinfo {volume} {122}},\ \bibinfo {pages}
  {241605} (\bibinfo {year} {2019})},\ \Eprint
  {http://arxiv.org/abs/1902.10571} {arXiv:1902.10571 [gr-qc]} \BibitemShut
  {NoStop}%
\bibitem [{\citenamefont {Bern}\ \emph
  {et~al.}(2019{\natexlab{c}})\citenamefont {Bern}, \citenamefont {Cheung},
  \citenamefont {Roiban}, \citenamefont {Shen}, \citenamefont {Solon},\ and\
  \citenamefont {Zeng}}]{Bern:2019crd}%
  \BibitemOpen
  \bibfield  {author} {\bibinfo {author} {\bibfnamefont {Z.}~\bibnamefont
  {Bern}}, \bibinfo {author} {\bibfnamefont {C.}~\bibnamefont {Cheung}},
  \bibinfo {author} {\bibfnamefont {R.}~\bibnamefont {Roiban}}, \bibinfo
  {author} {\bibfnamefont {C.-H.}\ \bibnamefont {Shen}}, \bibinfo {author}
  {\bibfnamefont {M.~P.}\ \bibnamefont {Solon}}, \ and\ \bibinfo {author}
  {\bibfnamefont {M.}~\bibnamefont {Zeng}},\ }\href {\doibase
  10.1007/JHEP10(2019)206} {\bibfield  {journal} {\bibinfo  {journal} {JHEP}\
  }\textbf {\bibinfo {volume} {10}},\ \bibinfo {pages} {206} (\bibinfo {year}
  {2019}{\natexlab{c}})},\ \Eprint {http://arxiv.org/abs/1908.01493}
  {arXiv:1908.01493 [hep-th]} \BibitemShut {NoStop}%
\bibitem [{\citenamefont {Cheung}\ and\ \citenamefont
  {Solon}(2020{\natexlab{a}})}]{Cheung:2020gyp}%
  \BibitemOpen
  \bibfield  {author} {\bibinfo {author} {\bibfnamefont {C.}~\bibnamefont
  {Cheung}}\ and\ \bibinfo {author} {\bibfnamefont {M.~P.}\ \bibnamefont
  {Solon}},\ }\href {\doibase 10.1007/JHEP06(2020)144} {\bibfield  {journal}
  {\bibinfo  {journal} {JHEP}\ }\textbf {\bibinfo {volume} {06}},\ \bibinfo
  {pages} {144} (\bibinfo {year} {2020}{\natexlab{a}})},\ \Eprint
  {http://arxiv.org/abs/2003.08351} {arXiv:2003.08351 [hep-th]} \BibitemShut
  {NoStop}%
\bibitem [{\citenamefont {K\"alin}\ and\ \citenamefont
  {Porto}(2020{\natexlab{a}})}]{Kalin:2020mvi}%
  \BibitemOpen
  \bibfield  {author} {\bibinfo {author} {\bibfnamefont {G.}~\bibnamefont
  {K\"alin}}\ and\ \bibinfo {author} {\bibfnamefont {R.~A.}\ \bibnamefont
  {Porto}},\ }\href {\doibase 10.1007/JHEP11(2020)106} {\bibfield  {journal}
  {\bibinfo  {journal} {JHEP}\ }\textbf {\bibinfo {volume} {11}},\ \bibinfo
  {pages} {106} (\bibinfo {year} {2020}{\natexlab{a}})},\ \Eprint
  {http://arxiv.org/abs/2006.01184} {arXiv:2006.01184 [hep-th]} \BibitemShut
  {NoStop}%
\bibitem [{\citenamefont {K\"alin}\ \emph
  {et~al.}(2020{\natexlab{a}})\citenamefont {K\"alin}, \citenamefont {Liu},\
  and\ \citenamefont {Porto}}]{Kalin:2020fhe}%
  \BibitemOpen
  \bibfield  {author} {\bibinfo {author} {\bibfnamefont {G.}~\bibnamefont
  {K\"alin}}, \bibinfo {author} {\bibfnamefont {Z.}~\bibnamefont {Liu}}, \ and\
  \bibinfo {author} {\bibfnamefont {R.~A.}\ \bibnamefont {Porto}},\ }\href@noop
  {} {\  (\bibinfo {year} {2020}{\natexlab{a}})},\ \Eprint
  {http://arxiv.org/abs/2007.04977} {arXiv:2007.04977 [hep-th]} \BibitemShut
  {NoStop}%
\bibitem [{\citenamefont {Cristofoli}\ \emph {et~al.}(2020)\citenamefont
  {Cristofoli}, \citenamefont {Damgaard}, \citenamefont {Di~Vecchia},\ and\
  \citenamefont {Heissenberg}}]{Cristofoli:2020uzm}%
  \BibitemOpen
  \bibfield  {author} {\bibinfo {author} {\bibfnamefont {A.}~\bibnamefont
  {Cristofoli}}, \bibinfo {author} {\bibfnamefont {P.~H.}\ \bibnamefont
  {Damgaard}}, \bibinfo {author} {\bibfnamefont {P.}~\bibnamefont
  {Di~Vecchia}}, \ and\ \bibinfo {author} {\bibfnamefont {C.}~\bibnamefont
  {Heissenberg}},\ }\href {\doibase 10.1007/JHEP07(2020)122} {\bibfield
  {journal} {\bibinfo  {journal} {JHEP}\ }\textbf {\bibinfo {volume} {07}},\
  \bibinfo {pages} {122} (\bibinfo {year} {2020})},\ \Eprint
  {http://arxiv.org/abs/2003.10274} {arXiv:2003.10274 [hep-th]} \BibitemShut
  {NoStop}%
\bibitem [{\citenamefont {Bini}\ \emph
  {et~al.}(2020{\natexlab{a}})\citenamefont {Bini}, \citenamefont {Damour},
  \citenamefont {Geralico}, \citenamefont {Laporta},\ and\ \citenamefont
  {Mastrolia}}]{Bini:2020uiq}%
  \BibitemOpen
  \bibfield  {author} {\bibinfo {author} {\bibfnamefont {D.}~\bibnamefont
  {Bini}}, \bibinfo {author} {\bibfnamefont {T.}~\bibnamefont {Damour}},
  \bibinfo {author} {\bibfnamefont {A.}~\bibnamefont {Geralico}}, \bibinfo
  {author} {\bibfnamefont {S.}~\bibnamefont {Laporta}}, \ and\ \bibinfo
  {author} {\bibfnamefont {P.}~\bibnamefont {Mastrolia}},\ }\href@noop {} {\
  (\bibinfo {year} {2020}{\natexlab{a}})},\ \Eprint
  {http://arxiv.org/abs/2008.09389} {arXiv:2008.09389 [gr-qc]} \BibitemShut
  {NoStop}%
\bibitem [{\citenamefont {Bini}\ \emph
  {et~al.}(2020{\natexlab{b}})\citenamefont {Bini}, \citenamefont {Damour},
  \citenamefont {Geralico}, \citenamefont {Laporta},\ and\ \citenamefont
  {Mastrolia}}]{Bini:2020rzn}%
  \BibitemOpen
  \bibfield  {author} {\bibinfo {author} {\bibfnamefont {D.}~\bibnamefont
  {Bini}}, \bibinfo {author} {\bibfnamefont {T.}~\bibnamefont {Damour}},
  \bibinfo {author} {\bibfnamefont {A.}~\bibnamefont {Geralico}}, \bibinfo
  {author} {\bibfnamefont {S.}~\bibnamefont {Laporta}}, \ and\ \bibinfo
  {author} {\bibfnamefont {P.}~\bibnamefont {Mastrolia}},\ }\href@noop {} {\
  (\bibinfo {year} {2020}{\natexlab{b}})},\ \Eprint
  {http://arxiv.org/abs/2012.12918} {arXiv:2012.12918 [gr-qc]} \BibitemShut
  {NoStop}%
\bibitem [{\citenamefont {Loebbert}\ \emph {et~al.}(2020)\citenamefont
  {Loebbert}, \citenamefont {Plefka}, \citenamefont {Shi},\ and\ \citenamefont
  {Wang}}]{Loebbert:2020aos}%
  \BibitemOpen
  \bibfield  {author} {\bibinfo {author} {\bibfnamefont {F.}~\bibnamefont
  {Loebbert}}, \bibinfo {author} {\bibfnamefont {J.}~\bibnamefont {Plefka}},
  \bibinfo {author} {\bibfnamefont {C.}~\bibnamefont {Shi}}, \ and\ \bibinfo
  {author} {\bibfnamefont {T.}~\bibnamefont {Wang}},\ }\href@noop {} {\
  (\bibinfo {year} {2020})},\ \Eprint {http://arxiv.org/abs/2012.14224}
  {arXiv:2012.14224 [hep-th]} \BibitemShut {NoStop}%
\bibitem [{\citenamefont {Bern}\ \emph {et~al.}(2021)\citenamefont {Bern},
  \citenamefont {Parra-Martinez}, \citenamefont {Roiban}, \citenamefont {Ruf},
  \citenamefont {Shen}, \citenamefont {Solon},\ and\ \citenamefont
  {Zeng}}]{Bern:2021dqo}%
  \BibitemOpen
  \bibfield  {author} {\bibinfo {author} {\bibfnamefont {Z.}~\bibnamefont
  {Bern}}, \bibinfo {author} {\bibfnamefont {J.}~\bibnamefont
  {Parra-Martinez}}, \bibinfo {author} {\bibfnamefont {R.}~\bibnamefont
  {Roiban}}, \bibinfo {author} {\bibfnamefont {M.~S.}\ \bibnamefont {Ruf}},
  \bibinfo {author} {\bibfnamefont {C.-H.}\ \bibnamefont {Shen}}, \bibinfo
  {author} {\bibfnamefont {M.~P.}\ \bibnamefont {Solon}}, \ and\ \bibinfo
  {author} {\bibfnamefont {M.}~\bibnamefont {Zeng}},\ }\href {\doibase
  10.1103/PhysRevLett.126.171601} {\bibfield  {journal} {\bibinfo  {journal}
  {Phys. Rev. Lett.}\ }\textbf {\bibinfo {volume} {126}},\ \bibinfo {pages}
  {171601} (\bibinfo {year} {2021})},\ \Eprint
  {http://arxiv.org/abs/2101.07254} {arXiv:2101.07254 [hep-th]} \BibitemShut
  {NoStop}%
\bibitem [{\citenamefont {Vaidya}(2015)}]{Vaidya:2014kza}%
  \BibitemOpen
  \bibfield  {author} {\bibinfo {author} {\bibfnamefont {V.}~\bibnamefont
  {Vaidya}},\ }\href {\doibase 10.1103/PhysRevD.91.024017} {\bibfield
  {journal} {\bibinfo  {journal} {Phys. Rev. D}\ }\textbf {\bibinfo {volume}
  {91}},\ \bibinfo {pages} {024017} (\bibinfo {year} {2015})},\ \Eprint
  {http://arxiv.org/abs/1410.5348} {arXiv:1410.5348 [hep-th]} \BibitemShut
  {NoStop}%
\bibitem [{\citenamefont {Vines}(2018)}]{Vines:2017hyw}%
  \BibitemOpen
  \bibfield  {author} {\bibinfo {author} {\bibfnamefont {J.}~\bibnamefont
  {Vines}},\ }\href {\doibase 10.1088/1361-6382/aaa3a8} {\bibfield  {journal}
  {\bibinfo  {journal} {Class. Quant. Grav.}\ }\textbf {\bibinfo {volume}
  {35}},\ \bibinfo {pages} {084002} (\bibinfo {year} {2018})},\ \Eprint
  {http://arxiv.org/abs/1709.06016} {arXiv:1709.06016 [gr-qc]} \BibitemShut
  {NoStop}%
\bibitem [{\citenamefont {Guevara}(2019)}]{Guevara:2017csg}%
  \BibitemOpen
  \bibfield  {author} {\bibinfo {author} {\bibfnamefont {A.}~\bibnamefont
  {Guevara}},\ }\href {\doibase 10.1007/JHEP04(2019)033} {\bibfield  {journal}
  {\bibinfo  {journal} {JHEP}\ }\textbf {\bibinfo {volume} {04}},\ \bibinfo
  {pages} {033} (\bibinfo {year} {2019})},\ \Eprint
  {http://arxiv.org/abs/1706.02314} {arXiv:1706.02314 [hep-th]} \BibitemShut
  {NoStop}%
\bibitem [{\citenamefont {Vines}\ \emph {et~al.}(2019)\citenamefont {Vines},
  \citenamefont {Steinhoff},\ and\ \citenamefont {Buonanno}}]{Vines:2018gqi}%
  \BibitemOpen
  \bibfield  {author} {\bibinfo {author} {\bibfnamefont {J.}~\bibnamefont
  {Vines}}, \bibinfo {author} {\bibfnamefont {J.}~\bibnamefont {Steinhoff}}, \
  and\ \bibinfo {author} {\bibfnamefont {A.}~\bibnamefont {Buonanno}},\ }\href
  {\doibase 10.1103/PhysRevD.99.064054} {\bibfield  {journal} {\bibinfo
  {journal} {Phys. Rev. D}\ }\textbf {\bibinfo {volume} {99}},\ \bibinfo
  {pages} {064054} (\bibinfo {year} {2019})},\ \Eprint
  {http://arxiv.org/abs/1812.00956} {arXiv:1812.00956 [gr-qc]} \BibitemShut
  {NoStop}%
\bibitem [{\citenamefont {Guevara}\ \emph
  {et~al.}(2019{\natexlab{a}})\citenamefont {Guevara}, \citenamefont
  {Ochirov},\ and\ \citenamefont {Vines}}]{Guevara:2018wpp}%
  \BibitemOpen
  \bibfield  {author} {\bibinfo {author} {\bibfnamefont {A.}~\bibnamefont
  {Guevara}}, \bibinfo {author} {\bibfnamefont {A.}~\bibnamefont {Ochirov}}, \
  and\ \bibinfo {author} {\bibfnamefont {J.}~\bibnamefont {Vines}},\ }\href
  {\doibase 10.1007/JHEP09(2019)056} {\bibfield  {journal} {\bibinfo  {journal}
  {JHEP}\ }\textbf {\bibinfo {volume} {09}},\ \bibinfo {pages} {056} (\bibinfo
  {year} {2019}{\natexlab{a}})},\ \Eprint {http://arxiv.org/abs/1812.06895}
  {arXiv:1812.06895 [hep-th]} \BibitemShut {NoStop}%
\bibitem [{\citenamefont {Chung}\ \emph {et~al.}(2019)\citenamefont {Chung},
  \citenamefont {Huang}, \citenamefont {Kim},\ and\ \citenamefont
  {Lee}}]{Chung:2018kqs}%
  \BibitemOpen
  \bibfield  {author} {\bibinfo {author} {\bibfnamefont {M.-Z.}\ \bibnamefont
  {Chung}}, \bibinfo {author} {\bibfnamefont {Y.-T.}\ \bibnamefont {Huang}},
  \bibinfo {author} {\bibfnamefont {J.-W.}\ \bibnamefont {Kim}}, \ and\
  \bibinfo {author} {\bibfnamefont {S.}~\bibnamefont {Lee}},\ }\href {\doibase
  10.1007/JHEP04(2019)156} {\bibfield  {journal} {\bibinfo  {journal} {JHEP}\
  }\textbf {\bibinfo {volume} {04}},\ \bibinfo {pages} {156} (\bibinfo {year}
  {2019})},\ \Eprint {http://arxiv.org/abs/1812.08752} {arXiv:1812.08752
  [hep-th]} \BibitemShut {NoStop}%
\bibitem [{\citenamefont {Guevara}\ \emph
  {et~al.}(2019{\natexlab{b}})\citenamefont {Guevara}, \citenamefont
  {Ochirov},\ and\ \citenamefont {Vines}}]{Guevara:2019fsj}%
  \BibitemOpen
  \bibfield  {author} {\bibinfo {author} {\bibfnamefont {A.}~\bibnamefont
  {Guevara}}, \bibinfo {author} {\bibfnamefont {A.}~\bibnamefont {Ochirov}}, \
  and\ \bibinfo {author} {\bibfnamefont {J.}~\bibnamefont {Vines}},\ }\href
  {\doibase 10.1103/PhysRevD.100.104024} {\bibfield  {journal} {\bibinfo
  {journal} {Phys. Rev. D}\ }\textbf {\bibinfo {volume} {100}},\ \bibinfo
  {pages} {104024} (\bibinfo {year} {2019}{\natexlab{b}})},\ \Eprint
  {http://arxiv.org/abs/1906.10071} {arXiv:1906.10071 [hep-th]} \BibitemShut
  {NoStop}%
\bibitem [{\citenamefont {Chung}\ \emph {et~al.}(2020)\citenamefont {Chung},
  \citenamefont {Huang},\ and\ \citenamefont {Kim}}]{Chung:2019duq}%
  \BibitemOpen
  \bibfield  {author} {\bibinfo {author} {\bibfnamefont {M.-Z.}\ \bibnamefont
  {Chung}}, \bibinfo {author} {\bibfnamefont {Y.-T.}\ \bibnamefont {Huang}}, \
  and\ \bibinfo {author} {\bibfnamefont {J.-W.}\ \bibnamefont {Kim}},\ }\href
  {\doibase 10.1007/JHEP09(2020)074} {\bibfield  {journal} {\bibinfo  {journal}
  {JHEP}\ }\textbf {\bibinfo {volume} {09}},\ \bibinfo {pages} {074} (\bibinfo
  {year} {2020})},\ \Eprint {http://arxiv.org/abs/1908.08463} {arXiv:1908.08463
  [hep-th]} \BibitemShut {NoStop}%
\bibitem [{\citenamefont {Damgaard}\ \emph {et~al.}(2019)\citenamefont
  {Damgaard}, \citenamefont {Haddad},\ and\ \citenamefont
  {Helset}}]{Damgaard:2019lfh}%
  \BibitemOpen
  \bibfield  {author} {\bibinfo {author} {\bibfnamefont {P.~H.}\ \bibnamefont
  {Damgaard}}, \bibinfo {author} {\bibfnamefont {K.}~\bibnamefont {Haddad}}, \
  and\ \bibinfo {author} {\bibfnamefont {A.}~\bibnamefont {Helset}},\ }\href
  {\doibase 10.1007/JHEP11(2019)070} {\bibfield  {journal} {\bibinfo  {journal}
  {JHEP}\ }\textbf {\bibinfo {volume} {11}},\ \bibinfo {pages} {070} (\bibinfo
  {year} {2019})},\ \Eprint {http://arxiv.org/abs/1908.10308} {arXiv:1908.10308
  [hep-ph]} \BibitemShut {NoStop}%
\bibitem [{\citenamefont {Aoude}\ \emph
  {et~al.}(2020{\natexlab{a}})\citenamefont {Aoude}, \citenamefont {Haddad},\
  and\ \citenamefont {Helset}}]{Aoude:2020onz}%
  \BibitemOpen
  \bibfield  {author} {\bibinfo {author} {\bibfnamefont {R.}~\bibnamefont
  {Aoude}}, \bibinfo {author} {\bibfnamefont {K.}~\bibnamefont {Haddad}}, \
  and\ \bibinfo {author} {\bibfnamefont {A.}~\bibnamefont {Helset}},\ }\href
  {\doibase 10.1007/JHEP05(2020)051} {\bibfield  {journal} {\bibinfo  {journal}
  {JHEP}\ }\textbf {\bibinfo {volume} {05}},\ \bibinfo {pages} {051} (\bibinfo
  {year} {2020}{\natexlab{a}})},\ \Eprint {http://arxiv.org/abs/2001.09164}
  {arXiv:2001.09164 [hep-th]} \BibitemShut {NoStop}%
\bibitem [{\citenamefont {Bern}\ \emph
  {et~al.}(2020{\natexlab{a}})\citenamefont {Bern}, \citenamefont {Luna},
  \citenamefont {Roiban}, \citenamefont {Shen},\ and\ \citenamefont
  {Zeng}}]{Bern:2020buy}%
  \BibitemOpen
  \bibfield  {author} {\bibinfo {author} {\bibfnamefont {Z.}~\bibnamefont
  {Bern}}, \bibinfo {author} {\bibfnamefont {A.}~\bibnamefont {Luna}}, \bibinfo
  {author} {\bibfnamefont {R.}~\bibnamefont {Roiban}}, \bibinfo {author}
  {\bibfnamefont {C.-H.}\ \bibnamefont {Shen}}, \ and\ \bibinfo {author}
  {\bibfnamefont {M.}~\bibnamefont {Zeng}},\ }\href@noop {} {\  (\bibinfo
  {year} {2020}{\natexlab{a}})},\ \Eprint {http://arxiv.org/abs/2005.03071}
  {arXiv:2005.03071 [hep-th]} \BibitemShut {NoStop}%
\bibitem [{\citenamefont {Guevara}\ \emph {et~al.}(2020)\citenamefont
  {Guevara}, \citenamefont {Maybee}, \citenamefont {Ochirov}, \citenamefont
  {O'Connell},\ and\ \citenamefont {Vines}}]{Guevara:2020xjx}%
  \BibitemOpen
  \bibfield  {author} {\bibinfo {author} {\bibfnamefont {A.}~\bibnamefont
  {Guevara}}, \bibinfo {author} {\bibfnamefont {B.}~\bibnamefont {Maybee}},
  \bibinfo {author} {\bibfnamefont {A.}~\bibnamefont {Ochirov}}, \bibinfo
  {author} {\bibfnamefont {D.}~\bibnamefont {O'Connell}}, \ and\ \bibinfo
  {author} {\bibfnamefont {J.}~\bibnamefont {Vines}},\ }\href@noop {} {\
  (\bibinfo {year} {2020})},\ \Eprint {http://arxiv.org/abs/2012.11570}
  {arXiv:2012.11570 [hep-th]} \BibitemShut {NoStop}%
\bibitem [{\citenamefont {Levi}\ \emph
  {et~al.}(2020{\natexlab{a}})\citenamefont {Levi}, \citenamefont {Mcleod},\
  and\ \citenamefont {Von~Hippel}}]{Levi:2020kvb}%
  \BibitemOpen
  \bibfield  {author} {\bibinfo {author} {\bibfnamefont {M.}~\bibnamefont
  {Levi}}, \bibinfo {author} {\bibfnamefont {A.~J.}\ \bibnamefont {Mcleod}}, \
  and\ \bibinfo {author} {\bibfnamefont {M.}~\bibnamefont {Von~Hippel}},\
  }\href@noop {} {\  (\bibinfo {year} {2020}{\natexlab{a}})},\ \Eprint
  {http://arxiv.org/abs/2003.02827} {arXiv:2003.02827 [hep-th]} \BibitemShut
  {NoStop}%
\bibitem [{\citenamefont {Levi}\ \emph
  {et~al.}(2020{\natexlab{b}})\citenamefont {Levi}, \citenamefont {Mcleod},\
  and\ \citenamefont {Von~Hippel}}]{Levi:2020uwu}%
  \BibitemOpen
  \bibfield  {author} {\bibinfo {author} {\bibfnamefont {M.}~\bibnamefont
  {Levi}}, \bibinfo {author} {\bibfnamefont {A.~J.}\ \bibnamefont {Mcleod}}, \
  and\ \bibinfo {author} {\bibfnamefont {M.}~\bibnamefont {Von~Hippel}},\
  }\href@noop {} {\  (\bibinfo {year} {2020}{\natexlab{b}})},\ \Eprint
  {http://arxiv.org/abs/2003.07890} {arXiv:2003.07890 [hep-th]} \BibitemShut
  {NoStop}%
\bibitem [{\citenamefont {Cheung}\ and\ \citenamefont
  {Solon}(2020{\natexlab{b}})}]{Cheung:2020sdj}%
  \BibitemOpen
  \bibfield  {author} {\bibinfo {author} {\bibfnamefont {C.}~\bibnamefont
  {Cheung}}\ and\ \bibinfo {author} {\bibfnamefont {M.~P.}\ \bibnamefont
  {Solon}},\ }\href {\doibase 10.1103/PhysRevLett.125.191601} {\bibfield
  {journal} {\bibinfo  {journal} {Phys. Rev. Lett.}\ }\textbf {\bibinfo
  {volume} {125}},\ \bibinfo {pages} {191601} (\bibinfo {year}
  {2020}{\natexlab{b}})},\ \Eprint {http://arxiv.org/abs/2006.06665}
  {arXiv:2006.06665 [hep-th]} \BibitemShut {NoStop}%
\bibitem [{\citenamefont {Haddad}\ and\ \citenamefont
  {Helset}(2020)}]{Haddad:2020que}%
  \BibitemOpen
  \bibfield  {author} {\bibinfo {author} {\bibfnamefont {K.}~\bibnamefont
  {Haddad}}\ and\ \bibinfo {author} {\bibfnamefont {A.}~\bibnamefont
  {Helset}},\ }\href {\doibase 10.1007/JHEP12(2020)024} {\bibfield  {journal}
  {\bibinfo  {journal} {JHEP}\ }\textbf {\bibinfo {volume} {12}},\ \bibinfo
  {pages} {024} (\bibinfo {year} {2020})},\ \Eprint
  {http://arxiv.org/abs/2008.04920} {arXiv:2008.04920 [hep-th]} \BibitemShut
  {NoStop}%
\bibitem [{\citenamefont {K\"alin}\ \emph
  {et~al.}(2020{\natexlab{b}})\citenamefont {K\"alin}, \citenamefont {Liu},\
  and\ \citenamefont {Porto}}]{Kalin:2020lmz}%
  \BibitemOpen
  \bibfield  {author} {\bibinfo {author} {\bibfnamefont {G.}~\bibnamefont
  {K\"alin}}, \bibinfo {author} {\bibfnamefont {Z.}~\bibnamefont {Liu}}, \ and\
  \bibinfo {author} {\bibfnamefont {R.~A.}\ \bibnamefont {Porto}},\ }\href
  {\doibase 10.1103/PhysRevD.102.124025} {\bibfield  {journal} {\bibinfo
  {journal} {Phys. Rev. D}\ }\textbf {\bibinfo {volume} {102}},\ \bibinfo
  {pages} {124025} (\bibinfo {year} {2020}{\natexlab{b}})},\ \Eprint
  {http://arxiv.org/abs/2008.06047} {arXiv:2008.06047 [hep-th]} \BibitemShut
  {NoStop}%
\bibitem [{\citenamefont {Brandhuber}\ and\ \citenamefont
  {Travaglini}(2020)}]{Brandhuber:2019qpg}%
  \BibitemOpen
  \bibfield  {author} {\bibinfo {author} {\bibfnamefont {A.}~\bibnamefont
  {Brandhuber}}\ and\ \bibinfo {author} {\bibfnamefont {G.}~\bibnamefont
  {Travaglini}},\ }\href {\doibase 10.1007/JHEP01(2020)010} {\bibfield
  {journal} {\bibinfo  {journal} {JHEP}\ }\textbf {\bibinfo {volume} {01}},\
  \bibinfo {pages} {010} (\bibinfo {year} {2020})},\ \Eprint
  {http://arxiv.org/abs/1905.05657} {arXiv:1905.05657 [hep-th]} \BibitemShut
  {NoStop}%
\bibitem [{\citenamefont {Accettulli~Huber}\ \emph
  {et~al.}(2020{\natexlab{a}})\citenamefont {Accettulli~Huber}, \citenamefont
  {Brandhuber}, \citenamefont {De~Angelis},\ and\ \citenamefont
  {Travaglini}}]{Huber:2019ugz}%
  \BibitemOpen
  \bibfield  {author} {\bibinfo {author} {\bibfnamefont {M.}~\bibnamefont
  {Accettulli~Huber}}, \bibinfo {author} {\bibfnamefont {A.}~\bibnamefont
  {Brandhuber}}, \bibinfo {author} {\bibfnamefont {S.}~\bibnamefont
  {De~Angelis}}, \ and\ \bibinfo {author} {\bibfnamefont {G.}~\bibnamefont
  {Travaglini}},\ }\href {\doibase 10.1103/PhysRevD.101.046011} {\bibfield
  {journal} {\bibinfo  {journal} {Phys. Rev. D}\ }\textbf {\bibinfo {volume}
  {101}},\ \bibinfo {pages} {046011} (\bibinfo {year} {2020}{\natexlab{a}})},\
  \Eprint {http://arxiv.org/abs/1911.10108} {arXiv:1911.10108 [hep-th]}
  \BibitemShut {NoStop}%
\bibitem [{\citenamefont {Accettulli~Huber}\ \emph
  {et~al.}(2020{\natexlab{b}})\citenamefont {Accettulli~Huber}, \citenamefont
  {Brandhuber}, \citenamefont {De~Angelis},\ and\ \citenamefont
  {Travaglini}}]{AccettulliHuber:2020oou}%
  \BibitemOpen
  \bibfield  {author} {\bibinfo {author} {\bibfnamefont {M.}~\bibnamefont
  {Accettulli~Huber}}, \bibinfo {author} {\bibfnamefont {A.}~\bibnamefont
  {Brandhuber}}, \bibinfo {author} {\bibfnamefont {S.}~\bibnamefont
  {De~Angelis}}, \ and\ \bibinfo {author} {\bibfnamefont {G.}~\bibnamefont
  {Travaglini}},\ }\href {\doibase 10.1103/PhysRevD.102.046014} {\bibfield
  {journal} {\bibinfo  {journal} {Phys. Rev. D}\ }\textbf {\bibinfo {volume}
  {102}},\ \bibinfo {pages} {046014} (\bibinfo {year} {2020}{\natexlab{b}})},\
  \Eprint {http://arxiv.org/abs/2006.02375} {arXiv:2006.02375 [hep-th]}
  \BibitemShut {NoStop}%
\bibitem [{\citenamefont {Bern}\ \emph
  {et~al.}(2020{\natexlab{b}})\citenamefont {Bern}, \citenamefont
  {Parra-Martinez}, \citenamefont {Roiban}, \citenamefont {Sawyer},\ and\
  \citenamefont {Shen}}]{Bern:2020uwk}%
  \BibitemOpen
  \bibfield  {author} {\bibinfo {author} {\bibfnamefont {Z.}~\bibnamefont
  {Bern}}, \bibinfo {author} {\bibfnamefont {J.}~\bibnamefont
  {Parra-Martinez}}, \bibinfo {author} {\bibfnamefont {R.}~\bibnamefont
  {Roiban}}, \bibinfo {author} {\bibfnamefont {E.}~\bibnamefont {Sawyer}}, \
  and\ \bibinfo {author} {\bibfnamefont {C.-H.}\ \bibnamefont {Shen}},\
  }\href@noop {} {\  (\bibinfo {year} {2020}{\natexlab{b}})},\ \Eprint
  {http://arxiv.org/abs/2010.08559} {arXiv:2010.08559 [hep-th]} \BibitemShut
  {NoStop}%
\bibitem [{\citenamefont {Cheung}\ \emph {et~al.}(2020)\citenamefont {Cheung},
  \citenamefont {Shah},\ and\ \citenamefont {Solon}}]{Cheung:2020gbf}%
  \BibitemOpen
  \bibfield  {author} {\bibinfo {author} {\bibfnamefont {C.}~\bibnamefont
  {Cheung}}, \bibinfo {author} {\bibfnamefont {N.}~\bibnamefont {Shah}}, \ and\
  \bibinfo {author} {\bibfnamefont {M.~P.}\ \bibnamefont {Solon}},\ }\href@noop
  {} {\  (\bibinfo {year} {2020})},\ \Eprint {http://arxiv.org/abs/2010.08568}
  {arXiv:2010.08568 [hep-th]} \BibitemShut {NoStop}%
\bibitem [{\citenamefont {Aoude}\ \emph
  {et~al.}(2020{\natexlab{b}})\citenamefont {Aoude}, \citenamefont {Haddad},\
  and\ \citenamefont {Helset}}]{Aoude:2020ygw}%
  \BibitemOpen
  \bibfield  {author} {\bibinfo {author} {\bibfnamefont {R.}~\bibnamefont
  {Aoude}}, \bibinfo {author} {\bibfnamefont {K.}~\bibnamefont {Haddad}}, \
  and\ \bibinfo {author} {\bibfnamefont {A.}~\bibnamefont {Helset}},\
  }\href@noop {} {\  (\bibinfo {year} {2020}{\natexlab{b}})},\ \Eprint
  {http://arxiv.org/abs/2012.05256} {arXiv:2012.05256 [hep-th]} \BibitemShut
  {NoStop}%
\bibitem [{\citenamefont {Cristofoli}\ \emph {et~al.}(2019)\citenamefont
  {Cristofoli}, \citenamefont {Bjerrum-Bohr}, \citenamefont {Damgaard},\ and\
  \citenamefont {Vanhove}}]{Cristofoli:2019neg}%
  \BibitemOpen
  \bibfield  {author} {\bibinfo {author} {\bibfnamefont {A.}~\bibnamefont
  {Cristofoli}}, \bibinfo {author} {\bibfnamefont {N.}~\bibnamefont
  {Bjerrum-Bohr}}, \bibinfo {author} {\bibfnamefont {P.~H.}\ \bibnamefont
  {Damgaard}}, \ and\ \bibinfo {author} {\bibfnamefont {P.}~\bibnamefont
  {Vanhove}},\ }\href {\doibase 10.1103/PhysRevD.100.084040} {\bibfield
  {journal} {\bibinfo  {journal} {Phys. Rev. D}\ }\textbf {\bibinfo {volume}
  {100}},\ \bibinfo {pages} {084040} (\bibinfo {year} {2019})},\ \Eprint
  {http://arxiv.org/abs/1906.01579} {arXiv:1906.01579 [hep-th]} \BibitemShut
  {NoStop}%
\bibitem [{\citenamefont {Goldberger}\ and\ \citenamefont
  {Ridgway}(2017)}]{Goldberger:2016iau}%
  \BibitemOpen
  \bibfield  {author} {\bibinfo {author} {\bibfnamefont {W.~D.}\ \bibnamefont
  {Goldberger}}\ and\ \bibinfo {author} {\bibfnamefont {A.~K.}\ \bibnamefont
  {Ridgway}},\ }\href {\doibase 10.1103/PhysRevD.95.125010} {\bibfield
  {journal} {\bibinfo  {journal} {Phys. Rev. D}\ }\textbf {\bibinfo {volume}
  {95}},\ \bibinfo {pages} {125010} (\bibinfo {year} {2017})},\ \Eprint
  {http://arxiv.org/abs/1611.03493} {arXiv:1611.03493 [hep-th]} \BibitemShut
  {NoStop}%
\bibitem [{\citenamefont {Luna}\ \emph {et~al.}(2018)\citenamefont {Luna},
  \citenamefont {Nicholson}, \citenamefont {O'Connell},\ and\ \citenamefont
  {White}}]{Luna:2017dtq}%
  \BibitemOpen
  \bibfield  {author} {\bibinfo {author} {\bibfnamefont {A.}~\bibnamefont
  {Luna}}, \bibinfo {author} {\bibfnamefont {I.}~\bibnamefont {Nicholson}},
  \bibinfo {author} {\bibfnamefont {D.}~\bibnamefont {O'Connell}}, \ and\
  \bibinfo {author} {\bibfnamefont {C.~D.}\ \bibnamefont {White}},\ }\href
  {\doibase 10.1007/JHEP03(2018)044} {\bibfield  {journal} {\bibinfo  {journal}
  {JHEP}\ }\textbf {\bibinfo {volume} {03}},\ \bibinfo {pages} {044} (\bibinfo
  {year} {2018})},\ \Eprint {http://arxiv.org/abs/1711.03901} {arXiv:1711.03901
  [hep-th]} \BibitemShut {NoStop}%
\bibitem [{\citenamefont {Shen}(2018)}]{Shen:2018ebu}%
  \BibitemOpen
  \bibfield  {author} {\bibinfo {author} {\bibfnamefont {C.-H.}\ \bibnamefont
  {Shen}},\ }\href {\doibase 10.1007/JHEP11(2018)162} {\bibfield  {journal}
  {\bibinfo  {journal} {JHEP}\ }\textbf {\bibinfo {volume} {11}},\ \bibinfo
  {pages} {162} (\bibinfo {year} {2018})},\ \Eprint
  {http://arxiv.org/abs/1806.07388} {arXiv:1806.07388 [hep-th]} \BibitemShut
  {NoStop}%
\bibitem [{\citenamefont {Bautista}\ and\ \citenamefont
  {Guevara}(2019)}]{Bautista:2019tdr}%
  \BibitemOpen
  \bibfield  {author} {\bibinfo {author} {\bibfnamefont {Y.~F.}\ \bibnamefont
  {Bautista}}\ and\ \bibinfo {author} {\bibfnamefont {A.}~\bibnamefont
  {Guevara}},\ }\href@noop {} {\  (\bibinfo {year} {2019})},\ \Eprint
  {http://arxiv.org/abs/1903.12419} {arXiv:1903.12419 [hep-th]} \BibitemShut
  {NoStop}%
\bibitem [{\citenamefont {Mogull}\ \emph {et~al.}(2020)\citenamefont {Mogull},
  \citenamefont {Plefka},\ and\ \citenamefont {Steinhoff}}]{Mogull:2020sak}%
  \BibitemOpen
  \bibfield  {author} {\bibinfo {author} {\bibfnamefont {G.}~\bibnamefont
  {Mogull}}, \bibinfo {author} {\bibfnamefont {J.}~\bibnamefont {Plefka}}, \
  and\ \bibinfo {author} {\bibfnamefont {J.}~\bibnamefont {Steinhoff}},\
  }\href@noop {} {\  (\bibinfo {year} {2020})},\ \Eprint
  {http://arxiv.org/abs/2010.02865} {arXiv:2010.02865 [hep-th]} \BibitemShut
  {NoStop}%
\bibitem [{\citenamefont {Accettulli~Huber}\ \emph
  {et~al.}(2020{\natexlab{c}})\citenamefont {Accettulli~Huber}, \citenamefont
  {Brandhuber}, \citenamefont {De~Angelis},\ and\ \citenamefont
  {Travaglini}}]{Huber:2020xny}%
  \BibitemOpen
  \bibfield  {author} {\bibinfo {author} {\bibfnamefont {M.}~\bibnamefont
  {Accettulli~Huber}}, \bibinfo {author} {\bibfnamefont {A.}~\bibnamefont
  {Brandhuber}}, \bibinfo {author} {\bibfnamefont {S.}~\bibnamefont
  {De~Angelis}}, \ and\ \bibinfo {author} {\bibfnamefont {G.}~\bibnamefont
  {Travaglini}},\ }\href@noop {} {\  (\bibinfo {year} {2020}{\natexlab{c}})},\
  \Eprint {http://arxiv.org/abs/2012.06548} {arXiv:2012.06548 [hep-th]}
  \BibitemShut {NoStop}%
\bibitem [{\citenamefont {K\"alin}\ and\ \citenamefont
  {Porto}(2020{\natexlab{b}})}]{Kalin:2019rwq}%
  \BibitemOpen
  \bibfield  {author} {\bibinfo {author} {\bibfnamefont {G.}~\bibnamefont
  {K\"alin}}\ and\ \bibinfo {author} {\bibfnamefont {R.~A.}\ \bibnamefont
  {Porto}},\ }\href {\doibase 10.1007/JHEP01(2020)072} {\bibfield  {journal}
  {\bibinfo  {journal} {JHEP}\ }\textbf {\bibinfo {volume} {01}},\ \bibinfo
  {pages} {072} (\bibinfo {year} {2020}{\natexlab{b}})},\ \Eprint
  {http://arxiv.org/abs/1910.03008} {arXiv:1910.03008 [hep-th]} \BibitemShut
  {NoStop}%
\bibitem [{\citenamefont {K\"alin}\ and\ \citenamefont
  {Porto}(2020{\natexlab{c}})}]{Kalin:2019inp}%
  \BibitemOpen
  \bibfield  {author} {\bibinfo {author} {\bibfnamefont {G.}~\bibnamefont
  {K\"alin}}\ and\ \bibinfo {author} {\bibfnamefont {R.~A.}\ \bibnamefont
  {Porto}},\ }\href {\doibase 10.1007/JHEP02(2020)120} {\bibfield  {journal}
  {\bibinfo  {journal} {JHEP}\ }\textbf {\bibinfo {volume} {02}},\ \bibinfo
  {pages} {120} (\bibinfo {year} {2020}{\natexlab{c}})},\ \Eprint
  {http://arxiv.org/abs/1911.09130} {arXiv:1911.09130 [hep-th]} \BibitemShut
  {NoStop}%
\bibitem [{\citenamefont {Bini}\ \emph
  {et~al.}(2020{\natexlab{c}})\citenamefont {Bini}, \citenamefont {Damour},\
  and\ \citenamefont {Geralico}}]{Bini:2020hmy}%
  \BibitemOpen
  \bibfield  {author} {\bibinfo {author} {\bibfnamefont {D.}~\bibnamefont
  {Bini}}, \bibinfo {author} {\bibfnamefont {T.}~\bibnamefont {Damour}}, \ and\
  \bibinfo {author} {\bibfnamefont {A.}~\bibnamefont {Geralico}},\ }\href
  {\doibase 10.1103/PhysRevD.102.084047} {\bibfield  {journal} {\bibinfo
  {journal} {Phys. Rev. D}\ }\textbf {\bibinfo {volume} {102}},\ \bibinfo
  {pages} {084047} (\bibinfo {year} {2020}{\natexlab{c}})},\ \Eprint
  {http://arxiv.org/abs/2007.11239} {arXiv:2007.11239 [gr-qc]} \BibitemShut
  {NoStop}%
\bibitem [{\citenamefont {Kosower}\ \emph {et~al.}(2019)\citenamefont
  {Kosower}, \citenamefont {Maybee},\ and\ \citenamefont
  {O'Connell}}]{Kosower:2018adc}%
  \BibitemOpen
  \bibfield  {author} {\bibinfo {author} {\bibfnamefont {D.~A.}\ \bibnamefont
  {Kosower}}, \bibinfo {author} {\bibfnamefont {B.}~\bibnamefont {Maybee}}, \
  and\ \bibinfo {author} {\bibfnamefont {D.}~\bibnamefont {O'Connell}},\ }\href
  {\doibase 10.1007/JHEP02(2019)137} {\bibfield  {journal} {\bibinfo  {journal}
  {JHEP}\ }\textbf {\bibinfo {volume} {02}},\ \bibinfo {pages} {137} (\bibinfo
  {year} {2019})},\ \Eprint {http://arxiv.org/abs/1811.10950} {arXiv:1811.10950
  [hep-th]} \BibitemShut {NoStop}%
\bibitem [{\citenamefont {A}\ \emph {et~al.}(2020)\citenamefont {A},
  \citenamefont {Ghosh}, \citenamefont {Laddha},\ and\ \citenamefont
  {Athira}}]{A:2020lub}%
  \BibitemOpen
  \bibfield  {author} {\bibinfo {author} {\bibfnamefont {M.}~\bibnamefont {A}},
  \bibinfo {author} {\bibfnamefont {D.}~\bibnamefont {Ghosh}}, \bibinfo
  {author} {\bibfnamefont {A.}~\bibnamefont {Laddha}}, \ and\ \bibinfo {author}
  {\bibfnamefont {P.}~\bibnamefont {Athira}},\ }\href@noop {} {\  (\bibinfo
  {year} {2020})},\ \Eprint {http://arxiv.org/abs/2007.02077} {arXiv:2007.02077
  [hep-th]} \BibitemShut {NoStop}%
\bibitem [{\citenamefont {Laddha}\ and\ \citenamefont
  {Sen}(2018{\natexlab{a}})}]{Laddha:2018rle}%
  \BibitemOpen
  \bibfield  {author} {\bibinfo {author} {\bibfnamefont {A.}~\bibnamefont
  {Laddha}}\ and\ \bibinfo {author} {\bibfnamefont {A.}~\bibnamefont {Sen}},\
  }\href {\doibase 10.1007/JHEP09(2018)105} {\bibfield  {journal} {\bibinfo
  {journal} {JHEP}\ }\textbf {\bibinfo {volume} {09}},\ \bibinfo {pages} {105}
  (\bibinfo {year} {2018}{\natexlab{a}})},\ \Eprint
  {http://arxiv.org/abs/1801.07719} {arXiv:1801.07719 [hep-th]} \BibitemShut
  {NoStop}%
\bibitem [{\citenamefont {Laddha}\ and\ \citenamefont
  {Sen}(2018{\natexlab{b}})}]{Laddha:2018myi}%
  \BibitemOpen
  \bibfield  {author} {\bibinfo {author} {\bibfnamefont {A.}~\bibnamefont
  {Laddha}}\ and\ \bibinfo {author} {\bibfnamefont {A.}~\bibnamefont {Sen}},\
  }\href {\doibase 10.1007/JHEP10(2018)056} {\bibfield  {journal} {\bibinfo
  {journal} {JHEP}\ }\textbf {\bibinfo {volume} {10}},\ \bibinfo {pages} {056}
  (\bibinfo {year} {2018}{\natexlab{b}})},\ \Eprint
  {http://arxiv.org/abs/1804.09193} {arXiv:1804.09193 [hep-th]} \BibitemShut
  {NoStop}%
\bibitem [{\citenamefont {Sahoo}\ and\ \citenamefont
  {Sen}(2019)}]{Sahoo:2018lxl}%
  \BibitemOpen
  \bibfield  {author} {\bibinfo {author} {\bibfnamefont {B.}~\bibnamefont
  {Sahoo}}\ and\ \bibinfo {author} {\bibfnamefont {A.}~\bibnamefont {Sen}},\
  }\href {\doibase 10.1007/JHEP02(2019)086} {\bibfield  {journal} {\bibinfo
  {journal} {JHEP}\ }\textbf {\bibinfo {volume} {02}},\ \bibinfo {pages} {086}
  (\bibinfo {year} {2019})},\ \Eprint {http://arxiv.org/abs/1808.03288}
  {arXiv:1808.03288 [hep-th]} \BibitemShut {NoStop}%
\bibitem [{\citenamefont {Laddha}\ and\ \citenamefont
  {Sen}(2020)}]{Laddha:2019yaj}%
  \BibitemOpen
  \bibfield  {author} {\bibinfo {author} {\bibfnamefont {A.}~\bibnamefont
  {Laddha}}\ and\ \bibinfo {author} {\bibfnamefont {A.}~\bibnamefont {Sen}},\
  }\href {\doibase 10.1103/PhysRevD.101.084011} {\bibfield  {journal} {\bibinfo
   {journal} {Phys. Rev. D}\ }\textbf {\bibinfo {volume} {101}},\ \bibinfo
  {pages} {084011} (\bibinfo {year} {2020})},\ \Eprint
  {http://arxiv.org/abs/1906.08288} {arXiv:1906.08288 [gr-qc]} \BibitemShut
  {NoStop}%
\bibitem [{\citenamefont {Saha}\ \emph {et~al.}(2020)\citenamefont {Saha},
  \citenamefont {Sahoo},\ and\ \citenamefont {Sen}}]{Saha:2019tub}%
  \BibitemOpen
  \bibfield  {author} {\bibinfo {author} {\bibfnamefont {A.~P.}\ \bibnamefont
  {Saha}}, \bibinfo {author} {\bibfnamefont {B.}~\bibnamefont {Sahoo}}, \ and\
  \bibinfo {author} {\bibfnamefont {A.}~\bibnamefont {Sen}},\ }\href {\doibase
  10.1007/JHEP06(2020)153} {\bibfield  {journal} {\bibinfo  {journal} {JHEP}\
  }\textbf {\bibinfo {volume} {06}},\ \bibinfo {pages} {153} (\bibinfo {year}
  {2020})},\ \Eprint {http://arxiv.org/abs/1912.06413} {arXiv:1912.06413
  [hep-th]} \BibitemShut {NoStop}%
\bibitem [{\citenamefont {Kotikov}(1991)}]{Kotikov:1990kg}%
  \BibitemOpen
  \bibfield  {author} {\bibinfo {author} {\bibfnamefont {A.}~\bibnamefont
  {Kotikov}},\ }\href {\doibase 10.1016/0370-2693(91)90413-K} {\bibfield
  {journal} {\bibinfo  {journal} {Phys. Lett. B}\ }\textbf {\bibinfo {volume}
  {254}},\ \bibinfo {pages} {158} (\bibinfo {year} {1991})}\BibitemShut
  {NoStop}%
\bibitem [{\citenamefont {Bern}\ \emph {et~al.}(1993)\citenamefont {Bern},
  \citenamefont {Dixon},\ and\ \citenamefont {Kosower}}]{Bern:1992em}%
  \BibitemOpen
  \bibfield  {author} {\bibinfo {author} {\bibfnamefont {Z.}~\bibnamefont
  {Bern}}, \bibinfo {author} {\bibfnamefont {L.~J.}\ \bibnamefont {Dixon}}, \
  and\ \bibinfo {author} {\bibfnamefont {D.~A.}\ \bibnamefont {Kosower}},\
  }\href {\doibase 10.1016/0370-2693(93)90400-C} {\bibfield  {journal}
  {\bibinfo  {journal} {Phys. Lett. B}\ }\textbf {\bibinfo {volume} {302}},\
  \bibinfo {pages} {299} (\bibinfo {year} {1993})},\ \bibinfo {note} {[Erratum:
  Phys.Lett.B 318, 649 (1993)]},\ \Eprint {http://arxiv.org/abs/hep-ph/9212308}
  {arXiv:hep-ph/9212308} \BibitemShut {NoStop}%
\bibitem [{\citenamefont {Gehrmann}\ and\ \citenamefont
  {Remiddi}(2000)}]{Gehrmann:1999as}%
  \BibitemOpen
  \bibfield  {author} {\bibinfo {author} {\bibfnamefont {T.}~\bibnamefont
  {Gehrmann}}\ and\ \bibinfo {author} {\bibfnamefont {E.}~\bibnamefont
  {Remiddi}},\ }\href {\doibase 10.1016/S0550-3213(00)00223-6} {\bibfield
  {journal} {\bibinfo  {journal} {Nucl. Phys. B}\ }\textbf {\bibinfo {volume}
  {580}},\ \bibinfo {pages} {485} (\bibinfo {year} {2000})},\ \Eprint
  {http://arxiv.org/abs/hep-ph/9912329} {arXiv:hep-ph/9912329} \BibitemShut
  {NoStop}%
\bibitem [{\citenamefont {Henn}(2013)}]{Henn:2013pwa}%
  \BibitemOpen
  \bibfield  {author} {\bibinfo {author} {\bibfnamefont {J.~M.}\ \bibnamefont
  {Henn}},\ }\href {\doibase 10.1103/PhysRevLett.110.251601} {\bibfield
  {journal} {\bibinfo  {journal} {Phys. Rev. Lett.}\ }\textbf {\bibinfo
  {volume} {110}},\ \bibinfo {pages} {251601} (\bibinfo {year} {2013})},\
  \Eprint {http://arxiv.org/abs/1304.1806} {arXiv:1304.1806 [hep-th]}
  \BibitemShut {NoStop}%
\bibitem [{\citenamefont {Henn}(2015)}]{Henn:2014qga}%
  \BibitemOpen
  \bibfield  {author} {\bibinfo {author} {\bibfnamefont {J.~M.}\ \bibnamefont
  {Henn}},\ }\href {\doibase 10.1088/1751-8113/48/15/153001} {\bibfield
  {journal} {\bibinfo  {journal} {J. Phys. A}\ }\textbf {\bibinfo {volume}
  {48}},\ \bibinfo {pages} {153001} (\bibinfo {year} {2015})},\ \Eprint
  {http://arxiv.org/abs/1412.2296} {arXiv:1412.2296 [hep-ph]} \BibitemShut
  {NoStop}%
\bibitem [{\citenamefont {Parra-Martinez}\ \emph {et~al.}(2020)\citenamefont
  {Parra-Martinez}, \citenamefont {Ruf},\ and\ \citenamefont
  {Zeng}}]{Parra-Martinez:2020dzs}%
  \BibitemOpen
  \bibfield  {author} {\bibinfo {author} {\bibfnamefont {J.}~\bibnamefont
  {Parra-Martinez}}, \bibinfo {author} {\bibfnamefont {M.~S.}\ \bibnamefont
  {Ruf}}, \ and\ \bibinfo {author} {\bibfnamefont {M.}~\bibnamefont {Zeng}},\
  }\href {\doibase 10.1007/JHEP11(2020)023} {\bibfield  {journal} {\bibinfo
  {journal} {JHEP}\ }\textbf {\bibinfo {volume} {11}},\ \bibinfo {pages} {023}
  (\bibinfo {year} {2020})},\ \Eprint {http://arxiv.org/abs/2005.04236}
  {arXiv:2005.04236 [hep-th]} \BibitemShut {NoStop}%
\bibitem [{\citenamefont {Anastasiou}\ and\ \citenamefont
  {Melnikov}(2002)}]{Anastasiou:2002yz}%
  \BibitemOpen
  \bibfield  {author} {\bibinfo {author} {\bibfnamefont {C.}~\bibnamefont
  {Anastasiou}}\ and\ \bibinfo {author} {\bibfnamefont {K.}~\bibnamefont
  {Melnikov}},\ }\href {\doibase 10.1016/S0550-3213(02)00837-4} {\bibfield
  {journal} {\bibinfo  {journal} {Nucl. Phys. B}\ }\textbf {\bibinfo {volume}
  {646}},\ \bibinfo {pages} {220} (\bibinfo {year} {2002})},\ \Eprint
  {http://arxiv.org/abs/hep-ph/0207004} {arXiv:hep-ph/0207004} \BibitemShut
  {NoStop}%
\bibitem [{\citenamefont {Anastasiou}\ \emph
  {et~al.}(2003{\natexlab{a}})\citenamefont {Anastasiou}, \citenamefont
  {Dixon},\ and\ \citenamefont {Melnikov}}]{Anastasiou:2002qz}%
  \BibitemOpen
  \bibfield  {author} {\bibinfo {author} {\bibfnamefont {C.}~\bibnamefont
  {Anastasiou}}, \bibinfo {author} {\bibfnamefont {L.~J.}\ \bibnamefont
  {Dixon}}, \ and\ \bibinfo {author} {\bibfnamefont {K.}~\bibnamefont
  {Melnikov}},\ }\href {\doibase 10.1016/S0920-5632(03)80168-8} {\bibfield
  {journal} {\bibinfo  {journal} {Nucl. Phys. B Proc. Suppl.}\ }\textbf
  {\bibinfo {volume} {116}},\ \bibinfo {pages} {193} (\bibinfo {year}
  {2003}{\natexlab{a}})},\ \Eprint {http://arxiv.org/abs/hep-ph/0211141}
  {arXiv:hep-ph/0211141} \BibitemShut {NoStop}%
\bibitem [{\citenamefont {Anastasiou}\ \emph
  {et~al.}(2003{\natexlab{b}})\citenamefont {Anastasiou}, \citenamefont
  {Dixon}, \citenamefont {Melnikov},\ and\ \citenamefont
  {Petriello}}]{Anastasiou:2003yy}%
  \BibitemOpen
  \bibfield  {author} {\bibinfo {author} {\bibfnamefont {C.}~\bibnamefont
  {Anastasiou}}, \bibinfo {author} {\bibfnamefont {L.~J.}\ \bibnamefont
  {Dixon}}, \bibinfo {author} {\bibfnamefont {K.}~\bibnamefont {Melnikov}}, \
  and\ \bibinfo {author} {\bibfnamefont {F.}~\bibnamefont {Petriello}},\ }\href
  {\doibase 10.1103/PhysRevLett.91.182002} {\bibfield  {journal} {\bibinfo
  {journal} {Phys. Rev. Lett.}\ }\textbf {\bibinfo {volume} {91}},\ \bibinfo
  {pages} {182002} (\bibinfo {year} {2003}{\natexlab{b}})},\ \Eprint
  {http://arxiv.org/abs/hep-ph/0306192} {arXiv:hep-ph/0306192} \BibitemShut
  {NoStop}%
\bibitem [{\citenamefont {Anastasiou}\ \emph {et~al.}(2015)\citenamefont
  {Anastasiou}, \citenamefont {Duhr}, \citenamefont {Dulat}, \citenamefont
  {Furlan}, \citenamefont {Herzog},\ and\ \citenamefont
  {Mistlberger}}]{Anastasiou:2015yha}%
  \BibitemOpen
  \bibfield  {author} {\bibinfo {author} {\bibfnamefont {C.}~\bibnamefont
  {Anastasiou}}, \bibinfo {author} {\bibfnamefont {C.}~\bibnamefont {Duhr}},
  \bibinfo {author} {\bibfnamefont {F.}~\bibnamefont {Dulat}}, \bibinfo
  {author} {\bibfnamefont {E.}~\bibnamefont {Furlan}}, \bibinfo {author}
  {\bibfnamefont {F.}~\bibnamefont {Herzog}}, \ and\ \bibinfo {author}
  {\bibfnamefont {B.}~\bibnamefont {Mistlberger}},\ }\href {\doibase
  10.1007/JHEP08(2015)051} {\bibfield  {journal} {\bibinfo  {journal} {JHEP}\
  }\textbf {\bibinfo {volume} {08}},\ \bibinfo {pages} {051} (\bibinfo {year}
  {2015})},\ \Eprint {http://arxiv.org/abs/1505.04110} {arXiv:1505.04110
  [hep-ph]} \BibitemShut {NoStop}%
\bibitem [{\citenamefont {Amati}\ \emph {et~al.}(1990)\citenamefont {Amati},
  \citenamefont {Ciafaloni},\ and\ \citenamefont {Veneziano}}]{Amati:1990xe}%
  \BibitemOpen
  \bibfield  {author} {\bibinfo {author} {\bibfnamefont {D.}~\bibnamefont
  {Amati}}, \bibinfo {author} {\bibfnamefont {M.}~\bibnamefont {Ciafaloni}}, \
  and\ \bibinfo {author} {\bibfnamefont {G.}~\bibnamefont {Veneziano}},\ }\href
  {\doibase 10.1016/0550-3213(90)90375-N} {\bibfield  {journal} {\bibinfo
  {journal} {Nucl. Phys. B}\ }\textbf {\bibinfo {volume} {347}},\ \bibinfo
  {pages} {550} (\bibinfo {year} {1990})}\BibitemShut {NoStop}%
\bibitem [{\citenamefont {Di~Vecchia}\ \emph {et~al.}(2019)\citenamefont
  {Di~Vecchia}, \citenamefont {Luna}, \citenamefont {Naculich}, \citenamefont
  {Russo}, \citenamefont {Veneziano},\ and\ \citenamefont
  {White}}]{DiVecchia:2019myk}%
  \BibitemOpen
  \bibfield  {author} {\bibinfo {author} {\bibfnamefont {P.}~\bibnamefont
  {Di~Vecchia}}, \bibinfo {author} {\bibfnamefont {A.}~\bibnamefont {Luna}},
  \bibinfo {author} {\bibfnamefont {S.~G.}\ \bibnamefont {Naculich}}, \bibinfo
  {author} {\bibfnamefont {R.}~\bibnamefont {Russo}}, \bibinfo {author}
  {\bibfnamefont {G.}~\bibnamefont {Veneziano}}, \ and\ \bibinfo {author}
  {\bibfnamefont {C.~D.}\ \bibnamefont {White}},\ }\href {\doibase
  10.1016/j.physletb.2019.134927} {\bibfield  {journal} {\bibinfo  {journal}
  {Phys. Lett. B}\ }\textbf {\bibinfo {volume} {798}},\ \bibinfo {pages}
  {134927} (\bibinfo {year} {2019})},\ \Eprint
  {http://arxiv.org/abs/1908.05603} {arXiv:1908.05603 [hep-th]} \BibitemShut
  {NoStop}%
\bibitem [{\citenamefont {Di~Vecchia}\ \emph
  {et~al.}(2020{\natexlab{a}})\citenamefont {Di~Vecchia}, \citenamefont
  {Naculich}, \citenamefont {Russo}, \citenamefont {Veneziano},\ and\
  \citenamefont {White}}]{DiVecchia:2019kta}%
  \BibitemOpen
  \bibfield  {author} {\bibinfo {author} {\bibfnamefont {P.}~\bibnamefont
  {Di~Vecchia}}, \bibinfo {author} {\bibfnamefont {S.~G.}\ \bibnamefont
  {Naculich}}, \bibinfo {author} {\bibfnamefont {R.}~\bibnamefont {Russo}},
  \bibinfo {author} {\bibfnamefont {G.}~\bibnamefont {Veneziano}}, \ and\
  \bibinfo {author} {\bibfnamefont {C.~D.}\ \bibnamefont {White}},\ }\href
  {\doibase 10.1007/JHEP03(2020)173} {\bibfield  {journal} {\bibinfo  {journal}
  {JHEP}\ }\textbf {\bibinfo {volume} {03}},\ \bibinfo {pages} {173} (\bibinfo
  {year} {2020}{\natexlab{a}})},\ \Eprint {http://arxiv.org/abs/1911.11716}
  {arXiv:1911.11716 [hep-th]} \BibitemShut {NoStop}%
\bibitem [{\citenamefont {Bern}\ \emph
  {et~al.}(2020{\natexlab{c}})\citenamefont {Bern}, \citenamefont {Ita},
  \citenamefont {Parra-Martinez},\ and\ \citenamefont {Ruf}}]{Bern:2020gjj}%
  \BibitemOpen
  \bibfield  {author} {\bibinfo {author} {\bibfnamefont {Z.}~\bibnamefont
  {Bern}}, \bibinfo {author} {\bibfnamefont {H.}~\bibnamefont {Ita}}, \bibinfo
  {author} {\bibfnamefont {J.}~\bibnamefont {Parra-Martinez}}, \ and\ \bibinfo
  {author} {\bibfnamefont {M.~S.}\ \bibnamefont {Ruf}},\ }\href {\doibase
  10.1103/PhysRevLett.125.031601} {\bibfield  {journal} {\bibinfo  {journal}
  {Phys. Rev. Lett.}\ }\textbf {\bibinfo {volume} {125}},\ \bibinfo {pages}
  {031601} (\bibinfo {year} {2020}{\natexlab{c}})},\ \Eprint
  {http://arxiv.org/abs/2002.02459} {arXiv:2002.02459 [hep-th]} \BibitemShut
  {NoStop}%
\bibitem [{\citenamefont {Di~Vecchia}\ \emph
  {et~al.}(2020{\natexlab{b}})\citenamefont {Di~Vecchia}, \citenamefont
  {Heissenberg}, \citenamefont {Russo},\ and\ \citenamefont
  {Veneziano}}]{DiVecchia:2020ymx}%
  \BibitemOpen
  \bibfield  {author} {\bibinfo {author} {\bibfnamefont {P.}~\bibnamefont
  {Di~Vecchia}}, \bibinfo {author} {\bibfnamefont {C.}~\bibnamefont
  {Heissenberg}}, \bibinfo {author} {\bibfnamefont {R.}~\bibnamefont {Russo}},
  \ and\ \bibinfo {author} {\bibfnamefont {G.}~\bibnamefont {Veneziano}},\
  }\href {\doibase 10.1016/j.physletb.2020.135924} {\bibfield  {journal}
  {\bibinfo  {journal} {Phys. Lett. B}\ }\textbf {\bibinfo {volume} {811}},\
  \bibinfo {pages} {135924} (\bibinfo {year} {2020}{\natexlab{b}})},\ \Eprint
  {http://arxiv.org/abs/2008.12743} {arXiv:2008.12743 [hep-th]} \BibitemShut
  {NoStop}%
\bibitem [{\citenamefont {Di~Vecchia}\ \emph
  {et~al.}(2021{\natexlab{a}})\citenamefont {Di~Vecchia}, \citenamefont
  {Heissenberg}, \citenamefont {Russo},\ and\ \citenamefont
  {Veneziano}}]{DiVecchia:2021ndb}%
  \BibitemOpen
  \bibfield  {author} {\bibinfo {author} {\bibfnamefont {P.}~\bibnamefont
  {Di~Vecchia}}, \bibinfo {author} {\bibfnamefont {C.}~\bibnamefont
  {Heissenberg}}, \bibinfo {author} {\bibfnamefont {R.}~\bibnamefont {Russo}},
  \ and\ \bibinfo {author} {\bibfnamefont {G.}~\bibnamefont {Veneziano}},\
  }\href@noop {} {\  (\bibinfo {year} {2021}{\natexlab{a}})},\ \Eprint
  {http://arxiv.org/abs/2101.05772} {arXiv:2101.05772 [hep-th]} \BibitemShut
  {NoStop}%
\bibitem [{\citenamefont {Damour}(2020)}]{Damour:2020tta}%
  \BibitemOpen
  \bibfield  {author} {\bibinfo {author} {\bibfnamefont {T.}~\bibnamefont
  {Damour}},\ }\href {\doibase 10.1103/PhysRevD.102.124008} {\bibfield
  {journal} {\bibinfo  {journal} {Phys. Rev. D}\ }\textbf {\bibinfo {volume}
  {102}},\ \bibinfo {pages} {124008} (\bibinfo {year} {2020})},\ \Eprint
  {http://arxiv.org/abs/2010.01641} {arXiv:2010.01641 [gr-qc]} \BibitemShut
  {NoStop}%
\bibitem [{\citenamefont {Kovacs}\ and\ \citenamefont
  {Thorne}(1978)}]{Kovacs:1978eu}%
  \BibitemOpen
  \bibfield  {author} {\bibinfo {author} {\bibfnamefont {S.}~\bibnamefont
  {Kovacs}}\ and\ \bibinfo {author} {\bibfnamefont {K.}~\bibnamefont
  {Thorne}},\ }\href {\doibase 10.1086/156350} {\bibfield  {journal} {\bibinfo
  {journal} {Astrophys. J.}\ }\textbf {\bibinfo {volume} {224}},\ \bibinfo
  {pages} {62} (\bibinfo {year} {1978})}\BibitemShut {NoStop}%
\bibitem [{\citenamefont {Blanchet}\ and\ \citenamefont
  {Schaefer}(1989)}]{Blanchet:1989cu}%
  \BibitemOpen
  \bibfield  {author} {\bibinfo {author} {\bibfnamefont {L.}~\bibnamefont
  {Blanchet}}\ and\ \bibinfo {author} {\bibfnamefont {G.}~\bibnamefont
  {Schaefer}},\ }\href@noop {} {\bibfield  {journal} {\bibinfo  {journal} {Mon.
  Not. Roy. Astron. Soc.}\ }\textbf {\bibinfo {volume} {239}},\ \bibinfo
  {pages} {845} (\bibinfo {year} {1989})},\ \bibinfo {note} {[Erratum:
  Mon.Not.Roy.Astron.Soc. 242, 704 (1990)]}\BibitemShut {NoStop}%
\bibitem [{\citenamefont {Peters}\ and\ \citenamefont
  {Mathews}(1963)}]{Peters:1963ux}%
  \BibitemOpen
  \bibfield  {author} {\bibinfo {author} {\bibfnamefont {P.}~\bibnamefont
  {Peters}}\ and\ \bibinfo {author} {\bibfnamefont {J.}~\bibnamefont
  {Mathews}},\ }\href {\doibase 10.1103/PhysRev.131.435} {\bibfield  {journal}
  {\bibinfo  {journal} {Phys. Rev.}\ }\textbf {\bibinfo {volume} {131}},\
  \bibinfo {pages} {435} (\bibinfo {year} {1963})}\BibitemShut {NoStop}%
\bibitem [{\citenamefont {Peters}(1964)}]{Peters:1964zz}%
  \BibitemOpen
  \bibfield  {author} {\bibinfo {author} {\bibfnamefont {P.}~\bibnamefont
  {Peters}},\ }\href {\doibase 10.1103/PhysRev.136.B1224} {\bibfield  {journal}
  {\bibinfo  {journal} {Phys. Rev.}\ }\textbf {\bibinfo {volume} {136}},\
  \bibinfo {pages} {B1224} (\bibinfo {year} {1964})}\BibitemShut {NoStop}%
\bibitem [{\citenamefont {Wagoner}\ and\ \citenamefont
  {Will}(1976)}]{Wagoner:1976am}%
  \BibitemOpen
  \bibfield  {author} {\bibinfo {author} {\bibfnamefont {R.}~\bibnamefont
  {Wagoner}}\ and\ \bibinfo {author} {\bibfnamefont {C.}~\bibnamefont {Will}},\
  }\href {\doibase 10.1086/154886} {\bibfield  {journal} {\bibinfo  {journal}
  {Astrophys. J.}\ }\textbf {\bibinfo {volume} {210}},\ \bibinfo {pages} {764}
  (\bibinfo {year} {1976})},\ \bibinfo {note} {[Erratum: Astrophys.J. 215, 984
  (1977)]}\BibitemShut {NoStop}%
\bibitem [{\citenamefont {Junker}\ and\ \citenamefont
  {Sch\"afer}(1992)}]{Junker:1992kle}%
  \BibitemOpen
  \bibfield  {author} {\bibinfo {author} {\bibfnamefont {W.}~\bibnamefont
  {Junker}}\ and\ \bibinfo {author} {\bibfnamefont {G.}~\bibnamefont
  {Sch\"afer}},\ }\href {\doibase 10.1093/mnras/254.1.146} {\bibfield
  {journal} {\bibinfo  {journal} {Mon. Not. Roy. Astron. Soc.}\ }\textbf
  {\bibinfo {volume} {254}},\ \bibinfo {pages} {146} (\bibinfo {year}
  {1992})}\BibitemShut {NoStop}%
\bibitem [{\citenamefont {Gopakumar}\ \emph {et~al.}(1997)\citenamefont
  {Gopakumar}, \citenamefont {Iyer},\ and\ \citenamefont
  {Iyer}}]{Gopakumar:1997ng}%
  \BibitemOpen
  \bibfield  {author} {\bibinfo {author} {\bibfnamefont {A.}~\bibnamefont
  {Gopakumar}}, \bibinfo {author} {\bibfnamefont {B.~R.}\ \bibnamefont {Iyer}},
  \ and\ \bibinfo {author} {\bibfnamefont {S.}~\bibnamefont {Iyer}},\ }\href
  {\doibase 10.1103/PhysRevD.57.6562} {\bibfield  {journal} {\bibinfo
  {journal} {Phys. Rev. D}\ }\textbf {\bibinfo {volume} {55}},\ \bibinfo
  {pages} {6030} (\bibinfo {year} {1997})},\ \bibinfo {note} {[Erratum:
  Phys.Rev.D 57, 6562 (1998)]},\ \Eprint {http://arxiv.org/abs/gr-qc/9703075}
  {arXiv:gr-qc/9703075} \BibitemShut {NoStop}%
\bibitem [{\citenamefont {Gopakumar}\ and\ \citenamefont
  {Iyer}(2002)}]{Gopakumar:2001dy}%
  \BibitemOpen
  \bibfield  {author} {\bibinfo {author} {\bibfnamefont {A.}~\bibnamefont
  {Gopakumar}}\ and\ \bibinfo {author} {\bibfnamefont {B.~R.}\ \bibnamefont
  {Iyer}},\ }\href {\doibase 10.1103/PhysRevD.65.084011} {\bibfield  {journal}
  {\bibinfo  {journal} {Phys. Rev. D}\ }\textbf {\bibinfo {volume} {65}},\
  \bibinfo {pages} {084011} (\bibinfo {year} {2002})},\ \Eprint
  {http://arxiv.org/abs/gr-qc/0110100} {arXiv:gr-qc/0110100} \BibitemShut
  {NoStop}%
\bibitem [{\citenamefont {Arun}\ \emph
  {et~al.}(2008{\natexlab{a}})\citenamefont {Arun}, \citenamefont {Blanchet},
  \citenamefont {Iyer},\ and\ \citenamefont {Qusailah}}]{Arun:2007sg}%
  \BibitemOpen
  \bibfield  {author} {\bibinfo {author} {\bibfnamefont {K.}~\bibnamefont
  {Arun}}, \bibinfo {author} {\bibfnamefont {L.}~\bibnamefont {Blanchet}},
  \bibinfo {author} {\bibfnamefont {B.~R.}\ \bibnamefont {Iyer}}, \ and\
  \bibinfo {author} {\bibfnamefont {M.~S.}\ \bibnamefont {Qusailah}},\ }\href
  {\doibase 10.1103/PhysRevD.77.064035} {\bibfield  {journal} {\bibinfo
  {journal} {Phys. Rev. D}\ }\textbf {\bibinfo {volume} {77}},\ \bibinfo
  {pages} {064035} (\bibinfo {year} {2008}{\natexlab{a}})},\ \Eprint
  {http://arxiv.org/abs/0711.0302} {arXiv:0711.0302 [gr-qc]} \BibitemShut
  {NoStop}%
\bibitem [{\citenamefont {Blanchet}(2014)}]{Blanchet:2013haa}%
  \BibitemOpen
  \bibfield  {author} {\bibinfo {author} {\bibfnamefont {L.}~\bibnamefont
  {Blanchet}},\ }\href {\doibase 10.12942/lrr-2014-2} {\bibfield  {journal}
  {\bibinfo  {journal} {Living Rev. Rel.}\ }\textbf {\bibinfo {volume} {17}},\
  \bibinfo {pages} {2} (\bibinfo {year} {2014})},\ \Eprint
  {http://arxiv.org/abs/1310.1528} {arXiv:1310.1528 [gr-qc]} \BibitemShut
  {NoStop}%
\bibitem [{\citenamefont {Beneke}\ and\ \citenamefont
  {Smirnov}(1998)}]{Beneke:1997zp}%
  \BibitemOpen
  \bibfield  {author} {\bibinfo {author} {\bibfnamefont {M.}~\bibnamefont
  {Beneke}}\ and\ \bibinfo {author} {\bibfnamefont {V.~A.}\ \bibnamefont
  {Smirnov}},\ }\href {\doibase 10.1016/S0550-3213(98)00138-2} {\bibfield
  {journal} {\bibinfo  {journal} {Nucl. Phys. B}\ }\textbf {\bibinfo {volume}
  {522}},\ \bibinfo {pages} {321} (\bibinfo {year} {1998})},\ \Eprint
  {http://arxiv.org/abs/hep-ph/9711391} {arXiv:hep-ph/9711391} \BibitemShut
  {NoStop}%
\bibitem [{\citenamefont {Misner}\ \emph {et~al.}(1973)\citenamefont {Misner},
  \citenamefont {Thorne},\ and\ \citenamefont {Wheeler}}]{Misner:1974qy}%
  \BibitemOpen
  \bibfield  {author} {\bibinfo {author} {\bibfnamefont {C.~W.}\ \bibnamefont
  {Misner}}, \bibinfo {author} {\bibfnamefont {K.}~\bibnamefont {Thorne}}, \
  and\ \bibinfo {author} {\bibfnamefont {J.}~\bibnamefont {Wheeler}},\
  }\href@noop {} {\emph {\bibinfo {title} {{Gravitation}}}}\ (\bibinfo
  {publisher} {W. H. Freeman},\ \bibinfo {address} {San Francisco},\ \bibinfo
  {year} {1973})\BibitemShut {NoStop}%
\bibitem [{\citenamefont {Koemans~Collado}\ \emph {et~al.}(2019)\citenamefont
  {Koemans~Collado}, \citenamefont {Di~Vecchia},\ and\ \citenamefont
  {Russo}}]{KoemansCollado:2019ggb}%
  \BibitemOpen
  \bibfield  {author} {\bibinfo {author} {\bibfnamefont {A.}~\bibnamefont
  {Koemans~Collado}}, \bibinfo {author} {\bibfnamefont {P.}~\bibnamefont
  {Di~Vecchia}}, \ and\ \bibinfo {author} {\bibfnamefont {R.}~\bibnamefont
  {Russo}},\ }\href {\doibase 10.1103/PhysRevD.100.066028} {\bibfield
  {journal} {\bibinfo  {journal} {Phys. Rev. D}\ }\textbf {\bibinfo {volume}
  {100}},\ \bibinfo {pages} {066028} (\bibinfo {year} {2019})},\ \Eprint
  {http://arxiv.org/abs/1904.02667} {arXiv:1904.02667 [hep-th]} \BibitemShut
  {NoStop}%
\bibitem [{\citenamefont {Kosmopoulos}(2020)}]{Kosmopoulos:2020pcd}%
  \BibitemOpen
  \bibfield  {author} {\bibinfo {author} {\bibfnamefont {D.}~\bibnamefont
  {Kosmopoulos}},\ }\href@noop {} {\  (\bibinfo {year} {2020})},\ \Eprint
  {http://arxiv.org/abs/2009.00141} {arXiv:2009.00141 [hep-th]} \BibitemShut
  {NoStop}%
\bibitem [{\citenamefont {Chetyrkin}\ and\ \citenamefont
  {Tkachov}(1981)}]{Chetyrkin:1981qh}%
  \BibitemOpen
  \bibfield  {author} {\bibinfo {author} {\bibfnamefont {K.}~\bibnamefont
  {Chetyrkin}}\ and\ \bibinfo {author} {\bibfnamefont {F.}~\bibnamefont
  {Tkachov}},\ }\href {\doibase 10.1016/0550-3213(81)90199-1} {\bibfield
  {journal} {\bibinfo  {journal} {Nucl. Phys. B}\ }\textbf {\bibinfo {volume}
  {192}},\ \bibinfo {pages} {159} (\bibinfo {year} {1981})}\BibitemShut
  {NoStop}%
\bibitem [{\citenamefont {Herrmann}\ \emph {et~al.}(2021)\citenamefont
  {Herrmann}, \citenamefont {Parra-Martinez}, \citenamefont {Ruf},\ and\
  \citenamefont {Zeng}}]{Herrmann:2021tct}%
  \BibitemOpen
  \bibfield  {author} {\bibinfo {author} {\bibfnamefont {E.}~\bibnamefont
  {Herrmann}}, \bibinfo {author} {\bibfnamefont {J.}~\bibnamefont
  {Parra-Martinez}}, \bibinfo {author} {\bibfnamefont {M.~S.}\ \bibnamefont
  {Ruf}}, \ and\ \bibinfo {author} {\bibfnamefont {M.}~\bibnamefont {Zeng}},\
  }\href@noop {} {\  (\bibinfo {year} {2021})},\ \Eprint
  {http://arxiv.org/abs/2104.03957} {arXiv:2104.03957 [hep-th]} \BibitemShut
  {NoStop}%
\bibitem [{\citenamefont {Sogaard}\ and\ \citenamefont
  {Zhang}(2014)}]{Sogaard:2014ila}%
  \BibitemOpen
  \bibfield  {author} {\bibinfo {author} {\bibfnamefont {M.}~\bibnamefont
  {Sogaard}}\ and\ \bibinfo {author} {\bibfnamefont {Y.}~\bibnamefont
  {Zhang}},\ }\href {\doibase 10.1007/JHEP07(2014)112} {\bibfield  {journal}
  {\bibinfo  {journal} {JHEP}\ }\textbf {\bibinfo {volume} {07}},\ \bibinfo
  {pages} {112} (\bibinfo {year} {2014})},\ \Eprint
  {http://arxiv.org/abs/1403.2463} {arXiv:1403.2463 [hep-th]} \BibitemShut
  {NoStop}%
\bibitem [{\citenamefont {Di~Vecchia}\ \emph
  {et~al.}(2021{\natexlab{b}})\citenamefont {Di~Vecchia}, \citenamefont
  {Heissenberg}, \citenamefont {Russo},\ and\ \citenamefont
  {Veneziano}}]{DiVecchia:2021bdo}%
  \BibitemOpen
  \bibfield  {author} {\bibinfo {author} {\bibfnamefont {P.}~\bibnamefont
  {Di~Vecchia}}, \bibinfo {author} {\bibfnamefont {C.}~\bibnamefont
  {Heissenberg}}, \bibinfo {author} {\bibfnamefont {R.}~\bibnamefont {Russo}},
  \ and\ \bibinfo {author} {\bibfnamefont {G.}~\bibnamefont {Veneziano}},\
  }\href@noop {} {\  (\bibinfo {year} {2021}{\natexlab{b}})},\ \Eprint
  {http://arxiv.org/abs/2104.03256} {arXiv:2104.03256 [hep-th]} \BibitemShut
  {NoStop}%
\bibitem [{\citenamefont {Blanchet}\ and\ \citenamefont
  {Schaefer}(1993)}]{Blanchet:1993ec}%
  \BibitemOpen
  \bibfield  {author} {\bibinfo {author} {\bibfnamefont {L.}~\bibnamefont
  {Blanchet}}\ and\ \bibinfo {author} {\bibfnamefont {G.}~\bibnamefont
  {Schaefer}},\ }\href {\doibase 10.1088/0264-9381/10/12/026} {\bibfield
  {journal} {\bibinfo  {journal} {Class. Quant. Grav.}\ }\textbf {\bibinfo
  {volume} {10}},\ \bibinfo {pages} {2699} (\bibinfo {year}
  {1993})}\BibitemShut {NoStop}%
\bibitem [{\citenamefont {Rieth}\ and\ \citenamefont
  {Schaefer}(1997)}]{Rieth:1997mk}%
  \BibitemOpen
  \bibfield  {author} {\bibinfo {author} {\bibfnamefont {R.}~\bibnamefont
  {Rieth}}\ and\ \bibinfo {author} {\bibfnamefont {G.}~\bibnamefont
  {Schaefer}},\ }\href {\doibase 10.1088/0264-9381/14/8/029} {\bibfield
  {journal} {\bibinfo  {journal} {Class. Quant. Grav.}\ }\textbf {\bibinfo
  {volume} {14}},\ \bibinfo {pages} {2357} (\bibinfo {year}
  {1997})}\BibitemShut {NoStop}%
\bibitem [{\citenamefont {Arun}\ \emph
  {et~al.}(2008{\natexlab{b}})\citenamefont {Arun}, \citenamefont {Blanchet},
  \citenamefont {Iyer},\ and\ \citenamefont {Qusailah}}]{Arun:2007rg}%
  \BibitemOpen
  \bibfield  {author} {\bibinfo {author} {\bibfnamefont {K.}~\bibnamefont
  {Arun}}, \bibinfo {author} {\bibfnamefont {L.}~\bibnamefont {Blanchet}},
  \bibinfo {author} {\bibfnamefont {B.~R.}\ \bibnamefont {Iyer}}, \ and\
  \bibinfo {author} {\bibfnamefont {M.~S.}\ \bibnamefont {Qusailah}},\ }\href
  {\doibase 10.1103/PhysRevD.77.064034} {\bibfield  {journal} {\bibinfo
  {journal} {Phys. Rev. D}\ }\textbf {\bibinfo {volume} {77}},\ \bibinfo
  {pages} {064034} (\bibinfo {year} {2008}{\natexlab{b}})},\ \Eprint
  {http://arxiv.org/abs/0711.0250} {arXiv:0711.0250 [gr-qc]} \BibitemShut
  {NoStop}%
\bibitem [{\citenamefont {Peters}(1970)}]{Peters:1970mx}%
  \BibitemOpen
  \bibfield  {author} {\bibinfo {author} {\bibfnamefont {P.}~\bibnamefont
  {Peters}},\ }\href {\doibase 10.1103/PhysRevD.1.1559} {\bibfield  {journal}
  {\bibinfo  {journal} {Phys. Rev. D}\ }\textbf {\bibinfo {volume} {1}},\
  \bibinfo {pages} {1559} (\bibinfo {year} {1970})}\BibitemShut {NoStop}%
\bibitem [{\citenamefont {Jakobsen}\ \emph {et~al.}(2021)\citenamefont
  {Jakobsen}, \citenamefont {Mogull}, \citenamefont {Plefka},\ and\
  \citenamefont {Steinhoff}}]{Jakobsen:2021smu}%
  \BibitemOpen
  \bibfield  {author} {\bibinfo {author} {\bibfnamefont {G.~U.}\ \bibnamefont
  {Jakobsen}}, \bibinfo {author} {\bibfnamefont {G.}~\bibnamefont {Mogull}},
  \bibinfo {author} {\bibfnamefont {J.}~\bibnamefont {Plefka}}, \ and\ \bibinfo
  {author} {\bibfnamefont {J.}~\bibnamefont {Steinhoff}},\ }\href@noop {} {\
  (\bibinfo {year} {2021})},\ \Eprint {http://arxiv.org/abs/2101.12688}
  {arXiv:2101.12688 [gr-qc]} \BibitemShut {NoStop}%
\bibitem [{\citenamefont {Mougiakakos}\ \emph {et~al.}(2021)\citenamefont
  {Mougiakakos}, \citenamefont {Riva},\ and\ \citenamefont
  {Vernizzi}}]{Mougiakakos:2021ckm}%
  \BibitemOpen
  \bibfield  {author} {\bibinfo {author} {\bibfnamefont {S.}~\bibnamefont
  {Mougiakakos}}, \bibinfo {author} {\bibfnamefont {M.~M.}\ \bibnamefont
  {Riva}}, \ and\ \bibinfo {author} {\bibfnamefont {F.}~\bibnamefont
  {Vernizzi}},\ }\href@noop {} {\  (\bibinfo {year} {2021})},\ \Eprint
  {http://arxiv.org/abs/2102.08339} {arXiv:2102.08339 [gr-qc]} \BibitemShut
  {NoStop}%
\bibitem [{\citenamefont {Gruzinov}\ and\ \citenamefont
  {Veneziano}(2016)}]{Gruzinov:2014moa}%
  \BibitemOpen
  \bibfield  {author} {\bibinfo {author} {\bibfnamefont {A.}~\bibnamefont
  {Gruzinov}}\ and\ \bibinfo {author} {\bibfnamefont {G.}~\bibnamefont
  {Veneziano}},\ }\href {\doibase 10.1088/0264-9381/33/12/125012} {\bibfield
  {journal} {\bibinfo  {journal} {Class. Quant. Grav.}\ }\textbf {\bibinfo
  {volume} {33}},\ \bibinfo {pages} {125012} (\bibinfo {year} {2016})},\
  \Eprint {http://arxiv.org/abs/1409.4555} {arXiv:1409.4555 [gr-qc]}
  \BibitemShut {NoStop}%
\bibitem [{\citenamefont {Ciafaloni}\ \emph {et~al.}(2016)\citenamefont
  {Ciafaloni}, \citenamefont {Colferai}, \citenamefont {Coradeschi},\ and\
  \citenamefont {Veneziano}}]{Ciafaloni:2015xsr}%
  \BibitemOpen
  \bibfield  {author} {\bibinfo {author} {\bibfnamefont {M.}~\bibnamefont
  {Ciafaloni}}, \bibinfo {author} {\bibfnamefont {D.}~\bibnamefont {Colferai}},
  \bibinfo {author} {\bibfnamefont {F.}~\bibnamefont {Coradeschi}}, \ and\
  \bibinfo {author} {\bibfnamefont {G.}~\bibnamefont {Veneziano}},\ }\href
  {\doibase 10.1103/PhysRevD.93.044052} {\bibfield  {journal} {\bibinfo
  {journal} {Phys. Rev. D}\ }\textbf {\bibinfo {volume} {93}},\ \bibinfo
  {pages} {044052} (\bibinfo {year} {2016})},\ \Eprint
  {http://arxiv.org/abs/1512.00281} {arXiv:1512.00281 [hep-th]} \BibitemShut
  {NoStop}%
\bibitem [{\citenamefont {Ciafaloni}\ \emph {et~al.}(2019)\citenamefont
  {Ciafaloni}, \citenamefont {Colferai},\ and\ \citenamefont
  {Veneziano}}]{Ciafaloni:2018uwe}%
  \BibitemOpen
  \bibfield  {author} {\bibinfo {author} {\bibfnamefont {M.}~\bibnamefont
  {Ciafaloni}}, \bibinfo {author} {\bibfnamefont {D.}~\bibnamefont {Colferai}},
  \ and\ \bibinfo {author} {\bibfnamefont {G.}~\bibnamefont {Veneziano}},\
  }\href {\doibase 10.1103/PhysRevD.99.066008} {\bibfield  {journal} {\bibinfo
  {journal} {Phys. Rev. D}\ }\textbf {\bibinfo {volume} {99}},\ \bibinfo
  {pages} {066008} (\bibinfo {year} {2019})},\ \Eprint
  {http://arxiv.org/abs/1812.08137} {arXiv:1812.08137 [hep-th]} \BibitemShut
  {NoStop}%
\bibitem [{\citenamefont {Bini}\ and\ \citenamefont
  {Damour}(2017)}]{Bini:2017wfr}%
  \BibitemOpen
  \bibfield  {author} {\bibinfo {author} {\bibfnamefont {D.}~\bibnamefont
  {Bini}}\ and\ \bibinfo {author} {\bibfnamefont {T.}~\bibnamefont {Damour}},\
  }\href {\doibase 10.1103/PhysRevD.96.064021} {\bibfield  {journal} {\bibinfo
  {journal} {Phys. Rev. D}\ }\textbf {\bibinfo {volume} {96}},\ \bibinfo
  {pages} {064021} (\bibinfo {year} {2017})},\ \Eprint
  {http://arxiv.org/abs/1706.06877} {arXiv:1706.06877 [gr-qc]} \BibitemShut
  {NoStop}%
\bibitem [{\citenamefont {Blanchet}\ \emph {et~al.}(2020)\citenamefont
  {Blanchet}, \citenamefont {Foffa}, \citenamefont {Larrouturou},\ and\
  \citenamefont {Sturani}}]{Blanchet:2019rjs}%
  \BibitemOpen
  \bibfield  {author} {\bibinfo {author} {\bibfnamefont {L.}~\bibnamefont
  {Blanchet}}, \bibinfo {author} {\bibfnamefont {S.}~\bibnamefont {Foffa}},
  \bibinfo {author} {\bibfnamefont {F.}~\bibnamefont {Larrouturou}}, \ and\
  \bibinfo {author} {\bibfnamefont {R.}~\bibnamefont {Sturani}},\ }\href
  {\doibase 10.1103/PhysRevD.101.084045} {\bibfield  {journal} {\bibinfo
  {journal} {Phys. Rev. D}\ }\textbf {\bibinfo {volume} {101}},\ \bibinfo
  {pages} {084045} (\bibinfo {year} {2020})},\ \Eprint
  {http://arxiv.org/abs/1912.12359} {arXiv:1912.12359 [gr-qc]} \BibitemShut
  {NoStop}%
\bibitem [{\citenamefont {Cremmer}\ and\ \citenamefont
  {Julia}(1979)}]{Cremmer:1979up}%
  \BibitemOpen
  \bibfield  {author} {\bibinfo {author} {\bibfnamefont {E.}~\bibnamefont
  {Cremmer}}\ and\ \bibinfo {author} {\bibfnamefont {B.}~\bibnamefont
  {Julia}},\ }\href {\doibase 10.1016/0550-3213(79)90331-6} {\bibfield
  {journal} {\bibinfo  {journal} {Nucl. Phys. B}\ }\textbf {\bibinfo {volume}
  {159}},\ \bibinfo {pages} {141} (\bibinfo {year} {1979})}\BibitemShut
  {NoStop}%
\bibitem [{\citenamefont {Caron-Huot}\ and\ \citenamefont
  {Zahraee}(2019)}]{Caron-Huot:2018ape}%
  \BibitemOpen
  \bibfield  {author} {\bibinfo {author} {\bibfnamefont {S.}~\bibnamefont
  {Caron-Huot}}\ and\ \bibinfo {author} {\bibfnamefont {Z.}~\bibnamefont
  {Zahraee}},\ }\href {\doibase 10.1007/JHEP07(2019)179} {\bibfield  {journal}
  {\bibinfo  {journal} {JHEP}\ }\textbf {\bibinfo {volume} {07}},\ \bibinfo
  {pages} {179} (\bibinfo {year} {2019})},\ \Eprint
  {http://arxiv.org/abs/1810.04694} {arXiv:1810.04694 [hep-th]} \BibitemShut
  {NoStop}%
\bibitem [{\citenamefont {Maybee}\ \emph {et~al.}(2019)\citenamefont {Maybee},
  \citenamefont {O'Connell},\ and\ \citenamefont {Vines}}]{Maybee:2019jus}%
  \BibitemOpen
  \bibfield  {author} {\bibinfo {author} {\bibfnamefont {B.}~\bibnamefont
  {Maybee}}, \bibinfo {author} {\bibfnamefont {D.}~\bibnamefont {O'Connell}}, \
  and\ \bibinfo {author} {\bibfnamefont {J.}~\bibnamefont {Vines}},\ }\href
  {\doibase 10.1007/JHEP12(2019)156} {\bibfield  {journal} {\bibinfo  {journal}
  {JHEP}\ }\textbf {\bibinfo {volume} {12}},\ \bibinfo {pages} {156} (\bibinfo
  {year} {2019})},\ \Eprint {http://arxiv.org/abs/1906.09260} {arXiv:1906.09260
  [hep-th]} \BibitemShut {NoStop}%
\bibitem [{\citenamefont {de~la Cruz}\ \emph {et~al.}(2020)\citenamefont {de~la
  Cruz}, \citenamefont {Maybee}, \citenamefont {O'Connell},\ and\ \citenamefont
  {Ross}}]{delaCruz:2020bbn}%
  \BibitemOpen
  \bibfield  {author} {\bibinfo {author} {\bibfnamefont {L.}~\bibnamefont
  {de~la Cruz}}, \bibinfo {author} {\bibfnamefont {B.}~\bibnamefont {Maybee}},
  \bibinfo {author} {\bibfnamefont {D.}~\bibnamefont {O'Connell}}, \ and\
  \bibinfo {author} {\bibfnamefont {A.}~\bibnamefont {Ross}},\ }\href {\doibase
  10.1007/JHEP12(2020)076} {\bibfield  {journal} {\bibinfo  {journal} {JHEP}\
  }\textbf {\bibinfo {volume} {12}},\ \bibinfo {pages} {076} (\bibinfo {year}
  {2020})},\ \Eprint {http://arxiv.org/abs/2009.03842} {arXiv:2009.03842
  [hep-th]} \BibitemShut {NoStop}%
\bibitem [{\citenamefont {Gonzo}\ and\ \citenamefont
  {Pokraka}(2020)}]{Gonzo:2020xza}%
  \BibitemOpen
  \bibfield  {author} {\bibinfo {author} {\bibfnamefont {R.}~\bibnamefont
  {Gonzo}}\ and\ \bibinfo {author} {\bibfnamefont {A.}~\bibnamefont
  {Pokraka}},\ }\href@noop {} {\  (\bibinfo {year} {2020})},\ \Eprint
  {http://arxiv.org/abs/2012.01406} {arXiv:2012.01406 [hep-th]} \BibitemShut
  {NoStop}%
\bibitem [{\citenamefont {Korchemsky}\ \emph {et~al.}(1997)\citenamefont
  {Korchemsky}, \citenamefont {Oderda},\ and\ \citenamefont
  {Sterman}}]{Korchemsky:1997sy}%
  \BibitemOpen
  \bibfield  {author} {\bibinfo {author} {\bibfnamefont {G.~P.}\ \bibnamefont
  {Korchemsky}}, \bibinfo {author} {\bibfnamefont {G.}~\bibnamefont {Oderda}},
  \ and\ \bibinfo {author} {\bibfnamefont {G.~F.}\ \bibnamefont {Sterman}},\
  }\href {\doibase 10.1063/1.53732} {\bibfield  {journal} {\bibinfo  {journal}
  {AIP Conf. Proc.}\ }\textbf {\bibinfo {volume} {407}},\ \bibinfo {pages}
  {988} (\bibinfo {year} {1997})},\ \Eprint
  {http://arxiv.org/abs/hep-ph/9708346} {arXiv:hep-ph/9708346} \BibitemShut
  {NoStop}%
\bibitem [{\citenamefont {Korchemsky}\ and\ \citenamefont
  {Sterman}(1999)}]{Korchemsky:1999kt}%
  \BibitemOpen
  \bibfield  {author} {\bibinfo {author} {\bibfnamefont {G.~P.}\ \bibnamefont
  {Korchemsky}}\ and\ \bibinfo {author} {\bibfnamefont {G.~F.}\ \bibnamefont
  {Sterman}},\ }\href {\doibase 10.1016/S0550-3213(99)00308-9} {\bibfield
  {journal} {\bibinfo  {journal} {Nucl. Phys. B}\ }\textbf {\bibinfo {volume}
  {555}},\ \bibinfo {pages} {335} (\bibinfo {year} {1999})},\ \Eprint
  {http://arxiv.org/abs/hep-ph/9902341} {arXiv:hep-ph/9902341} \BibitemShut
  {NoStop}%
\end{thebibliography}%

\end{document}